\documentstyle[11pt,amssymb]{article}

\makeatletter
\@addtoreset{equation}{section}
\makeatother

\def\dalemb#1#2{{\vbox{\hrule height .#2pt
        \hbox{\vrule width.#2pt height#1pt \kern#1pt
                \vrule width.#2pt}
        \hrule height.#2pt}}}

\def\ca{{\cal I}}
\def\caj{{\cal J}}
\def\caP{{\cal P}}

\def\0{{\sst{(0)}}}
\def\1{{\sst{(1)}}}
\def\2{{\sst{(2)}}}
\def\3{{\sst{(3)}}}
\def\4{{\sst{(4)}}}
\def\5{{\sst{(5)}}}
\def\6{{\sst{(6)}}}
\def\7{{\sst{(7)}}}
\def\8{{\sst{(8)}}}

\def\ep{\epsilon}
\def\td{\tilde}

\def\half{{\textstyle{1\over2}}}
\def\qu{{\textstyle{1\over 4}}}

\def\eig{{\textstyle{1\over 8}}}
\let\a=\alpha \let\b=\beta \let\g=\gamma \let\d=\delta \let\e=\epsilon
  \let\q=\theta  \let\k=\kappa
\let\l=\lambda \let\m=\mu   
\let\s=\sigma \let\t=\tau  \let\f=\phi  \let\y=\psi
\let\w=\omega  \let\D=\Delta  
    
 \let\W=\Omega   \let\G=\Gamma
\let\la=\label  
   
\def\nn{\nonumber} \def\bd{\begin{document}} \def\ed{\end{document}}
\def\ds{\documentstyle} \let\fr=\frac \let\bl=\bigl \let\br=\bigr
\let\Br=\Bigr \let\Bl=\Bigl 
\let\bm=\bibitem
\let\na=\nabla
\let\pa=\partial \let\ov=\overline 
\newcommand{\be}{\begin{equation}} 
\newcommand{\ee}{\end{equation}}
\def\ba{\begin{array}}
\def\ea{\end{array}}
\def\ft#1#2{{\textstyle{{\scriptstyle #1}\over {\scriptstyle #2}}}}
\def\fft#1#2{{#1 \over #2}}
\def\del{\partial}
\def\sst#1{{\scriptscriptstyle #1}}
 \def\oneone{\rlap 1\mkern4mu{\rm l}}
\def\ie{{\it i.e.\ }}
\def\via{{\it via}}
\def\semi{{\ltimes}}
\def\str{{\rm str}}
\def\Dm{{{D_{\sst{max}}}}}
\def\vac{ \left | 0 \right \rangle }
\def\kvac{ \left | k \right \rangle }
\def\lb {\left .}
\def\rb {\right |}
\def\sp{\; \; \;}

\newcommand{\hsp}{\hspace{0.5cm}}

\newcommand{\ho}[1]{$\, ^{#1}$}
\newcommand{\hoch}[1]{$\, ^{#1}$}
\newcommand{\bea}{\begin{eqnarray}} 
\newcommand{\eea}{\end{eqnarray}} 
\newcommand{\ra}{\rightarrow}
\newcommand{\lra}{\longrightarrow}
\newcommand{\Lra}{\Leftrightarrow}
\newcommand{\ap}{\alpha^\prime}
\newcommand{\bp}{\tilde \beta^\prime}
\newcommand{\tr}{{\rm tr} }
\newcommand{\Tr}{{\rm Tr} } 
\newcommand{\NP}{Nucl. Phys. }
\newcommand{\damtp} {\it DAMTP, University of Cambridge, 
CMS, Wilberforce Road, Cambridge. CB3 0WE}
\newcommand{\spin}{{\it Spinoza Institute, University of Utrecht,\\
Postbus 80.195, 3508 TD Utrecht, The Netherlands}\\
{\tt email:taylor@phys.uu.nl}}
\newcommand{\princeton}{{\it Physics Department, Princeton University,\\ 
Princeton, NJ 08544, USA}\\
{\tt email:kostas@feynman.princeton.edu}}
\newcommand{\auth}{\large\bf{ 
Kostas Skenderis\hoch{\star} and Marika Taylor\hoch{\dagger}}}

\begin{document}

\begin{flushright}
\hfill{\bf hep-th/0211011}\\
\hfill{SPIN-2002/31} \\
\hfill{ITF-2002/51} \\
\hfill{PUPT-2056}
\end{flushright}

\vspace{15pt}

\begin{center}

{\Large \bf Open strings in the plane wave background I: \\
Quantization and symmetries}

\vspace{20pt}

\auth
\vspace{15pt}

{\hoch\star \princeton}

\vspace{8pt}

{\hoch\dagger \spin}

\vspace{15pt}

\underline{ABSTRACT}
\end{center}
We systematically investigate open strings in the  
plane wave background of type IIB string theory.
We carefully analyze possible boundary conditions for open
strings and find static as well as time-dependent branes.
The branes fall into equivalence classes depending on 
whether they are related by the action of target space
isometries. In particular static branes localized at the 
origin of transverse space 
and certain time-dependent branes 
fall into the same equivalence class. We analyze thoroughly
the symmetries of all branes we discuss. Apart from 
symmetries descending from target space isometries,
the worldsheet action being free admits a 
countably infinite number of other global worldsheet symmetries. 
We find that one can use such worldsheet symmetries to restore
seemingly broken target space symmetries.
In particular, we show that D-branes localized at 
arbitrary constant positions which were thought to be 1/4 
supersymmetric in fact have sixteen supercharges whilst 
D-branes which 
were thought to be non-supersymmetric have eight supercharges.
We discuss in detail the quantization in all cases.

\noindent

\pagebreak
\setcounter{page}{1}

\tableofcontents
\addtocontents{toc}{\protect\setcounter{tocdepth}{2}}

\newpage

\section{Introduction and summary of results}
\noindent

Understanding string theory on backgrounds with Ramond-Ramond (RR) fluxes 
is an important subject. Apart from its intrinsic importance,
it has acquired a renewed interest due to the AdS/CFT duality. 
Understanding the gauge theory/gravity correspondence in the regime 
where the gauge theory is weakly coupled requires understanding
string theory on such backgrounds.
Despite considerable effort, however, our understanding 
remains somewhat limited.  

Progress was made recently in \cite{Met,mt}
where it was shown that closed string theory on a specific 
plane wave background with a RR flux
is exactly solvable in the light cone gauge. 
This background is particularly 
interesting because it is obtained from the $AdS_5 \times S^5$ 
solution by a Penrose limit \cite{blau0,blau}. 
The AdS/CFT correspondence then suggests a direct relation 
between gauge theory and string theory on this background, 
and indeed such a connection was proposed in \cite{BMN}.

Given the importance of string theory on RR backgrounds,
it is of obvious interest to understand open strings and 
D-branes on such backgrounds. D-branes capture non-perturbative
phenomena in string theory, and as such are bound to 
play an important role in the gravity/gauge theory
correspondence. Moreover, much of our intuition 
about D-branes comes from flat space computations,
and understanding in detail cases where the
background is curved and contains RR fields
may provide us with new insights.

Motivated by these issues, we start in this paper a systematic 
investigation of open strings and D-branes
on the maximally supersymmetric plane 
wave background of type IIB supergravity \cite{blau0},
henceforth called ``the plane wave background''. 
We analyze in this paper the possible D-branes,
the quantization of open strings 
and the symmetries of the open string action.
In a companion paper we will discuss the corresponding superalgebras 
and the spectra of these D-branes \cite{ST3}.

D-branes on the plane wave background have been investigated by 
various authors. Boundary states were constructed in \cite{billo,BGG}, 
D-branes in open string theory were studied in \cite{DP, BPZ}, 
supersymmetric embeddings were classified in \cite{ST},
supergravity solutions were constructed in \cite{BMZ},
open-closed string duality was addressed in \cite{BGG} 
and the duality with gauge theory was discussed in 
\cite{Lee:2002cu,BMN2,ST,BHLN}. Aspects of D-branes on this and 
other plane wave backgrounds were also discussed in \cite{TT, brane}.
All these investigations yielded results
consistent with each other. We extend these results in 
various directions and in particular find that certain D-branes 
are more supersymmetric than found previously. 
We will show that certain worldsheet symmetries can be 
used to restored seemingly broken target space symmetries.
The additional supersymmetries originate from such a mechanism.
  
Let us first review the existing results. Recall that in
the plane wave background the eight directions transverse to the
lightcone are divided into two sets of four. 
The target space supersymmetry in the lightcone gauge can be 
divided into dynamical supersymmetries $Q^-$ and kinematical
supersymmetries  $Q^+$. We shall call ``dynamical'' the supercharges
that anticommute to the Hamiltonian (plus possibly other generators)
and ``kinematical'' the supercharges that anticommute to 
$P^+$. In the lightcone gauge the latter is just a c-number.
The kinematical supercharges when evaluated on-shell 
are proportional to zero modes, but the dynamical ones 
depend on all modes.  The half-supersymmetric
D-branes are found to be of the type $(+,-,m,m \pm 2)$,
where the notation indicates the worldvolume coordinates
and $m$ ($m \pm 2$) is the number of worldvolume coordinates 
in the first (second) set of four transverse directions. 
These D-branes preserve eight dynamical and eight 
kinematical supersymmetries when 
located at the origin of transverse space \cite{DP,ST}, 
but only eight kinematical ones when located at an arbitrary
constant position \cite{ST}. D-branes
with different splitting of coordinates were found to break
supersymmetry completely except for the case of the D1 brane
which preserves eight dynamical supersymmetries irrespectively of its
transverse position. 
Supersymmetric embeddings for D-branes that are not visible in open 
string theory in the 
light-cone gauge (because they wrap only one of the 
light-cone directions) were also discussed in \cite{ST}.

The plane wave/gauge theory duality for the D5 $(+,-,3,1)$ case
was discussed in \cite{Lee:2002cu}, for the D7 $(+,-,4,2)$
in \cite{BMN2} and a uniform discussion for all D-branes 
was presented in \cite{ST}.
In particular, it was proposed, and evidence was given, 
that all half supersymmetric D-branes are dual to defect conformal 
field theories. It was shown in \cite{ST} that the
$(+,-,m,m \pm 2)$ probe branes located in the 
origin of transverse space 
originate from $AdS_{m+1} \times S^{m+1 \pm 2}$
embeddings in $AdS_5 \times S^5$ after the Penrose limit. 
The latter embeddings originate from orthogonal intersections 
$D(2m+1\pm2) \perp D3$ in the 
near-horizon limit. The defect conformal field
theories involved in the gauge theory/plane wave duality
are the theories of the 
$(2m+2\pm2)$-$3$ strings (i.e. the strings that stretch between
the $D(2m+1\pm2)$ and the $D3$ brane in the orthogonal 
intersection $D(2m+1\pm2) \perp D3$). 

In this paper we revisit the analysis of open string theory
in the plane wave background. Such an analysis for (some) static
D-branes was presented in \cite{DP,BPZ} and we will comment below 
on the relation of our results to theirs. 
The first step in our analysis is to determine the 
possible boundary conditions consistent with the variational problem.
%
In addition, the Green-Schwarz action should be invariant under 
local $\k$-symmetry
and the closed string action constructed in \cite{Met} is indeed
$\k$-symmetric. In order for the corresponding open string to be invariant 
under $\k$-symmetry one has to verify that the boundary terms
that arise when checking the invariance of the action also vanish.
Such a computation was carried out in \cite{LW} for flat space, 
and in \cite{BPZ} for the plane wave background. From the results
of \cite{BPZ} one finds that when one imposes the 
fermionic lightcone gauge $\bar{\g}^+ \q =0$ the
boundary terms for all the branes we consider vanish. 
In \cite{BPZ} it was also shown 
that the boundary terms vanish for branes of type  
$(+,-,m,m \pm 2)$ localized at the origin
of transverse space without imposing $\bar{\g}^+ \q =0$.
In our case this condition is always satisfied:
the starting point of our analysis is the action where the kappa 
symmetry has already been gauge fixed, and we soon impose the 
lightcone and conformal gauges as well. 

In our analysis we discuss both time-dependent as well as static
branes. The static D-branes
are the direct analogue of the ordinary flat space D-branes, i.e.
they involve Neumann boundary conditions along the worldvolume of the brane
and Dirichlet boundary conditions along the transverse directions.
However, as reviewed above, 
the branes are also distinguished by the distribution
of worldvolume coordinates in the transverse to the lightcone 
directions. It is convenient to introduce the notation $D_-$ 
for the branes with worldvolume
coordinates  $(+,-,m,m \pm 2)$ and $D_+$ for all other $(+,-,m,n)$ branes
(recall that in open string theory in the lightcone 
gauge, the lightcone directions are necessarily worldvolume coordinates).
The reason for this terminology will become apparent in next section.
Notice that there are only $D_-p$ branes for $p=3,5,7$.

The boundary conditions corresponding to static branes break some
of the target space isometries. One may use the broken
generators to generate new branes which we call here
``symmetry-related branes''. These are equivalent 
to the original ones since the two are related by an 
isometric reparametrization of the target spacetime.
In particular, we discuss in detail the branes obtained
by translations and/or boosts of the static branes that are localized
at the origin of transverse space. These symmetry-related branes
satisfy time-dependent boundary conditions and in certain cases
describe Penrose limits of giant gravitons and other rotating branes.
 
The boundary conditions are time dependent since in Brinkmann
coordinates, which are the natural ones for string
quantization, the corresponding Killing vectors depend on $x^+$ \cite{blau0} 
and $x^+$ is identified with the worldsheet time $\tau$ 
in the light-cone gauge. As emphasized in \cite{BMZ},
such branes are static in the Rosen coordinate system, that is, their 
transverse position is constant, since here the translations
act as in flat space \cite{blau}. Notice that the symmetry-related
branes that are located away from the origin are 
distinct from static branes located at arbitrary
constant position. (The latter in turn would translate to time-dependent
branes in the Rosen coordinate system.) We also discuss more general 
time-dependent boundary conditions that are compatible 
with the variational problem.

We then proceed to quantize the open string with the
boundary conditions just discussed. Since the model
is free quantization is straightforward:
one first expands in modes and then imposes the 
canonical commutation relations. The $D_-$ 
and $D_+$ branes differ in their fermionic
zero modes. For $D_-$ branes these zero modes
depend only on the worldsheet time coordinate $\tau$, see (\ref{fer}),
and can be obtained from the zero modes of the 
closed string fermions, see (\ref{fercl}), upon identification. 
In the case of $D_+$ branes, however, 
the zero modes of the fermions depend on the worldsheet space coordinate
$\s$, see (\ref{fer2}), and there is no direct relation with the 
zero modes of the closed string. 

The commutation relations for $D_-$ branes were discussed 
in \cite{DP}, and for type $D1$-branes in \cite{BPZ}. Our  
results differ from those in these papers 
in numerical coefficients of the fermionic anticommutators,
and in that the anticommutator of the zero modes for all $D_+$ branes
is non-trivial. Note that our results have the correct flat space limit
including coefficients.
  
The target space isometries that are compatible with the boundary 
conditions become symmetries of the open string action.
We discuss in detail these symmetries and
compute the corresponding Noether currents
for all cases. In particular, part of the target space
supersymmetries, which from now on will be called 
``closed string supersymmetries'', are compatible with the 
open string boundary conditions. Recall that for every 
Killing spinor of the background $\k$-symmetry 
implies that the closed string worldsheet action is invariant 
under a target space global supersymmetry. In the open string case, 
one has to take into account the boundary conditions. 
This was done in \cite{BPZ} and it was found that the 
allowed D-branes preserve exactly the target space supersymmetries 
found in the analysis of \cite{ST}. 
 
One of our main results is that one can restore some of the broken 
symmetries by using worldsheet symmetries.
The worldsheet theory in the light cone gauge is quadratic in the 
fields. This implies that that there are an infinite number of 
symmetries. The transformation rules are
given by shifting each field by a parameter that satisfies 
the field equations. Such transformations leave the Lagrangian invariant up 
to total derivative terms. The transformations that are 
compatible with the boundary conditions are true 
symmetries of the theory. Expanding the parameter in a 
basis we find that there are a countably infinite number of 
such symmetries. The corresponding charges 
when evaluated on-shell are proportional to oscillators.

Worldsheet transformations that are not compatible with the boundary 
conditions play a role as well. One may combine them with a 
seemingly broken target space symmetry to obtain a 
good symmetry. In this way we shall show that we can
find eight additional dynamical supersymmetries in the case
of $D_-$ branes located away from the origin, and eight 
additional kinematical supersymmetries in the case of D1
branes. Thus these branes have sixteen supersymmetries but
the additional supercharges do not originate
from closed string supercharges. 

One may understand why the branes possess 
additional supercharges as follows. In flat space,
open string boundary conditions compatible with $2 N$ target
space supercharges lead to $N$ fermionic
oscillators at the $k$th mode. In the plane wave case
the number of closed string supersymmetries preserved
by our branes is in some cases less than 16, but
there are still $8$ fermionic oscillators
at each mode level. These oscillators are spectrum generating
and in all cases the spectrum automatically falls 
into supersymmetry representations.
The corresponding supercharges, however, do 
not descend from target space superisometries, but 
are instead the new supercharges.

The paper is organized as follows. In the next section we 
discuss the variational problem for the open string 
in the plane wave background and possible consistent
boundary conditions (D-branes). 
In section 3 we discuss the quantization of the open 
string for all cases considered in section 2
and in section 4 the corresponding symmetries 
and conserved charges.  We conclude in section 5
with a discussion of our results. We include several appendices.
In appendix A we give conventions and in 
appendices B and C we review results from closed
strings in the plane wave background. In appendix D 
we discuss in detail the computation of the 
(anti)commutation relation for the oscillators.
Appendix E contains a computation of the supersymmetry
of certain time dependent D-branes from the D-brane worldvolume
perspective.

\section{Action and boundary conditions}

In what follows we will consider open strings propagating in the 
maximally supersymmetric plane wave background \cite{blau0} with 
Brinkmann metric
\be  \label{ppwvmet}
ds^2 = 2 dx^{+} dx^{-} + \sum_{I=1}^{8} 
( dx^{I} dx^{I} - \mu^2 (x^I)^2 (dx^{+})^2),
\ee
and RR flux
\be 
F_{+1234} = F_{+5678} = 4 \mu.
\ee
It is useful to retain explicitly the mass parameter $\mu$ rather
than scale it to one since the flat space limit will then be 
manifest\footnote{ 
We use the supergravity conventions given in appendix A of 
\cite{ST}. Compared with the conventions in \cite{Met} 
one should rescale $F_5$ by a factor of 2. 
We also note that the coordinates used 
here are related to those used in (6.12)-(6.13) of \cite{ST} by 
$x^+ \to \mu x^+, x^- \to - 1/(2 \mu) x^{-}, x^I \to x^I$.}.

The Green-Schwarz action in this background was constructed by Metsaev
using supercoset techniques \cite{Met}. 
Just as for flat space, if one fixes the
lightcone fermionic gauge the action becomes quadratic in fermions:
the $\kappa$ symmetry gauge fixed Green-Schwarz action 
takes the form
\bea
S &=& T \int_\Sigma d^2\s \left ( -\half \sqrt{-g} g^{ab}
(2 \del_{a} x^{+} \del_{b} x^{-} - \mu^2 x_{I}^2 \del_{a} x^{+} \del_{b} x^{+}
+ \del_{a} x^{I} \del_{b} x^{I}) \right .\nn \\
&& \hsp - i 
\sqrt{-g} g^{ab} \del_{b} x^{+} (\bar{\q} \bar{\g}^{-} \del_{a} \q
+ \q \bar{\g}^{-} \del_{a} \bar{\q} + 2 i \mu \del_{a} x^{+} \bar{\q} 
\bar{\g}^{-} \Pi \q ) \label{kfix} \\
&& \hsp \left . + i 
\ep^{ab} \del_{a} x^{+} (\q \bar{\g}^{-} \del_{b} \q + 
\bar{\q} \bar{\g}^{-} \del_{b} \bar{\q}) \right ). \nn
\eea
In this expression $g_{ab}$ is the worldsheet metric with $(\t,\s)$
the worldsheet coordinates and $\ep^{01} =1$. $(x^{+},x^{-},x^{I})$ 
are the bosonic coordinates of the target superspace 
whilst $\q$ is a complex Weyl spinor
satisfying the fermionic lightcone gauge condition $\bar{\g}^{+} \q = 0$.  
Here we use the spinor notation of \cite{Met}, namely
$\g^{\mu}$ and $\bar{\g}^{\mu}$ are the the off-diagonal
parts of the 32-dimensional gamma matrices $\G^{\mu}$. Our conventions
are given in appendix A. Note that the complex Weyl spinor can be
replaced by two real Majorana spinors such that 
\be
\q = \frac{1}{\sqrt{2}} (\q^1 + i \q^2);
\ee
we will use both in what follows, according to convenience.
The explicit appearance of $\Pi = \g^{1234}$ in the action reflects the 
fact that the RR flux background breaks the transverse 
symmetry to $SO(4) \times SO(4)'$. The overall normalization
of the action $T$ is taken to be the inverse of the length of the
string (that is, $2 \pi$ for the closed string and $\pi$ for the
open string). 

\bigskip

We would now like to derive the field equations for the open
strings. To do this one should
consider variations of the action (\ref{kfix}); these will include
both bulk and boundary terms which may schematically be represented
as
\be
\d S = T \int_\Sigma d^2 \sigma \delta z^{m} F_{m} (z^{n}) + 
\int_{\pa \Sigma} d \sigma_{a} 
\delta z^{m} G_{m}^{a} (z^{n}),
\ee
where $z^{m}$ extends over all the coordinates of the target superspace,
which in this case are $(x^{+},x^{-},x^{I},\q^1,\q^2)$,
and the second integral is over the worldsheet boundary, $\pa \Sigma$. One
should then choose boundary conditions and if necessary add boundary
terms such that the variational problem is well defined, i.e. so
that $\delta S = 0$ implies the field equations. We discuss
this procedure in detail here since we are interested in non-standard
boundary conditions. 

Usually the boundary conditions chosen will automatically set to
zero $\delta z^{m} G_{m}^{a}$ for each $m$, and then $F_{m} = 0$ will give
us the full complement of field equations. However, one may also wish
to implement boundary conditions for which one or more component
of $\delta z^m G_{m}^{a}$ is non-vanishing and is not cancelled by
another component. In this case it would 
be necessary to add further boundary terms to the action 
such that under variations
with these boundary conditions the total variation is zero on
the boundary (for each $m$) {\footnote {One would also have
to check that corresponding boundary terms added to the original
action (before fixing kappa symmetry) ensure that the total
action is $\kappa$ symmetric with these boundary conditions.}}. 
We will impose
fermionic boundary conditions such that $\delta \q^1 G^{\q^1}_{a}$
cancels with $\delta \q^2 G^{\q^2}_a$; this relates $\q^1$ to $\q^2$ 
on the boundary and reduces the number of independent fermionic
degrees of freedom propagating on the worldsheet. Note that locality
requires that the boundary terms vanish separately at each boundary:
only non-local conditions could implement cancellation between terms
at different boundaries.

Before presenting the explicit forms for the variations of the action
(\ref{kfix}) let us comment that by starting with an action
in which the kappa symmetry has already been fixed we will necessarily
exclude boundary conditions (and hence D-branes) which are not compatible
with this fermionic gauge choice. These include branes along one
lightcone direction discussed in \cite{ST}. Furthermore, as
mentioned in the introduction, it is important that the 
boundary conditions we consider here are, using the results of \cite{BPZ},  
compatible with the kappa symmetry of the original
non gauge-fixed action given in \cite{Met}.

\bigskip

So let us now consider the variation of the bulk action (\ref{kfix}).
Under $x^{-}$ variations one obtains
\be
\d S = T \left(\int_\Sigma d^2 \s \d x^{-} \del_a - \int_{\pa \Sigma} 
d\s_a \d x^{-} \right) \sqrt{-g} g^{ab} \del_{b} x^{+},
\label{v1}
\ee
whilst under $x^{I}$ variations one finds
\bea
\d S &=& T \int_\Sigma d^2 \s 
\left( \del_{a} (\sqrt{-g} g^{ab} \del_{b} x^{I}) + 
\mu^2 x^{I}  \sqrt{-g} g^{ab} \del_{a} x^{+} \del_{b} x^{+} \right) 
\d x^{I} \label{v2} \\
& & - T \int_{\pa \Sigma} d\s_{a} \sqrt{-g} g^{ab} \del_{b} x^{I} \d x^{I}. \nn
\eea
The $x^{+}$ variations give
\bea
\d S &=& T \left(\int_\Sigma d^2 \s \d x^{+} \del_a - \int_{\pa \Sigma} 
d\s_a \d x^{+} \right) 
\left(- i \ep^{ab} (\q^1 \bar{\g}^{-} 
\del_{b} \q^1 - \q^2 \bar{\g}^{-} \del_{b} \q^2) +
\right.  \la{v3} \\
&& \left.
\sqrt{-g} g^{ab} [\del_{b} x^{-} - \mu^2 x_{I}^2 \del_{b} x^{+} 
+ i (\q^1 \bar{\g}^{-} 
\del_{b} \q^1 + \q^2 \bar{\g}^{-} \del_{b} \q^2
- 4 \mu \del_{b} x^{+} \q^1 \bar{\g}^{-} \Pi \q^2) ] \right). \nn 
\eea
The $\q^1$ variations give
\bea
\d S &=& T \int_\Sigma d^2 \s \d \q^1    
\left ( 
- i \del_{a} (\sqrt{g} g^{ab} 
\del_{b} x^+) (\bar{\g}^{-} \q^1) \right . \label{v4} \\
&& \left . 
- 2 i \sqrt{-g} g^{ab} \del_{b} x^{+} (\bar{\g}^{-} \del_{a} \q^1
-\mu \del_{a} x^{+} \bar{\g}^{-} \Pi \q^2) 
+ 2 i \ep^{ab} \del_{a} x^{+} 
\bar{\g}^{-} \del_{b} \q^1 \right ) \nn \\
&& + T \int_{\pa \Sigma} d \s_{a} \left ( \sqrt{-g} g^{ab} 
 + \ep^{ab} \right) i \del_{b} x^{+} \d \q^1 \bar{\g}^{-} \q^1  \nn
\eea
whilst the $\q^2$ variations give
\bea
\d S &=&  T \int_\Sigma d^2 \s \d \q^2    
\left ( 
- i \del_{a} (\sqrt{g} g^{ab} 
\del_{b} x^+) (\bar{\g}^{-} \q^2) \right . \label{v5} \\
&& \left . 
- 2 i \sqrt{-g} g^{ab} \del_{b} x^{+} (\bar{\g}^{-} \del_{a} \q^2
+\mu \del_{a} x^{+} \bar{\g}^{-} \Pi \q^1) 
- 2 i \ep^{ab} \del_{a} x^{+} 
\bar{\g}^{-} \del_{b} \q^1 \right ) \nn \\
&& + T \int_{\pa \Sigma} d \s_{a} \left ( \sqrt{-g} g^{ab} 
 - \ep^{ab} \right) i \del_{b} x^{+} \d \q^2 \bar{\g}^{-} \q^2 . \nn
\eea
Furthermore the worldsheet energy momentum tensor is
\bea
T_{ab} &=& -{2 \over T \sqrt{-g}} {\d S \over \d g^{ab}}=
L_{ab} - \half (g^{cd} L_{cd}) g_{ab}; \\
L_{ab} &=& 2 \del_{(a} x^{+} \del_{b)} x^{-} 
- \mu^2 x_I^2 \del_a x^+ \del_b x^+ + \del_a x^I \del_b x^I \nn \\
&& + 2 i \del_{(a} x^{+} ( \q^1 \bar{\g}^{-} 
\del_{b)} \q^1 + \q^2 \bar{\g}^{-} \del_{b)} \q^2 - 2 \mu 
\del_{b)} x^{+} \q^1 \bar{\g}^{-} \Pi \q^2 ), \label{V1}
\eea
Here and in
all that follows we (anti)symmetrise with unit strength. 
Notice that the stress energy tensor is traceless, i.e.
\be
g^{ab} T_{ab} =0.
\ee
The field equations obtained by varying the 
worldsheet metric (Virasoro constraints) amount to $T_{ab} = 0$. 

\bigskip

In the following sections we will consider open strings with 
a variety of different boundary conditions correspondent to D-branes.
We now describe in detail these boundary conditions.

\subsection{Static D-branes}

The simplest and most
familiar choice of boundary conditions corresponds to static D-branes. 
For these, the bosonic boundary conditions are as follows.
We first impose Neumann conditions on $(p-1)$ transverse
coordinates and 
Dirichlet boundary conditions on the remaining transverse coordinates:
\bea
\del_{\s} x^{r} | &=& 0, \hsp r = 1,..,(p-1) \label{bc1} \\
\del_{\t} x^{r'} | &=& 0, \hsp r' = p,..,8. \nn
\eea
Our index conventions are as follows: we use indices, 
$r,s,t=1,..,(p-1)$, 
from the end of  the latin alphabet
to denote coordinates with Neumann boundary conditions, and 
primed indices of the same letters, $r',s',t'=p,..,8$, to denote 
coordinates with Dirichlet boundary conditions. Latin 
indices, $i,j=1,..,4$, and $i',j'=5,..,8$, 
from the middle of the alphabet are used for the two sets 
of the transverse coordinates that transform among themselves
under the $SO(4)$ and the $SO(4)'$, respectively.

We fix the boundary of the worldsheet to be at $\s = 0,\pi$ and
denote by $A {|}$ the restriction of a quantity to the boundary.
We also at this stage fix conformal gauge for the worldsheet metric
$- g_{00} = g_{11} = 1$. 
Note that as usual we have assumed that all variations vanish
at $\tau \rightarrow \pm \infty$ and so we need only consider the
spatial boundary of the worldsheet. We also choose $x^{+}$ 
to satisfy a Neumann condition.

Now such boundary conditions remove all boundary terms in 
(\ref{v1}) and (\ref{v2}). The resulting field equation for
$x^+$ together with the Neumann boundary condition forces
us into bosonic lightcone gauge $x^+ = p^+ \t$. (Note that
one can also include an overall integration constant $x^+_0$
but since it does not play a role we do not include it here.)

\bigskip

Appropriate fermion boundary conditions are
\be
\q^1|  = \Omega \q^2{|}, \label{bc2}
\ee
at $\s=0$ and $\s=\pi$. 
$\Omega$ is a real (constant) $16 \times 16$ dimensional matrix.
Different matrices $\W$ can be used at $\s=0$ and $\s = \pi$, as 
representing open string extending between distinct branes, 
but here we take $\Omega$ to be the same at both ends.
Such a boundary condition necessarily reduces the number of independent
fermion modes by one half and preserves at most one half of the
supersymmetry. 
As previously advertised, the fermionic boundary condition (\ref{bc2})
can cancel boundary terms between (\ref{v4}) and
(\ref{v5}). That is, given the Neumann boundary
condition on $x^+$ and since $\d \q^1 = \Omega \d \q^2$ on the boundary
to preserve (\ref{bc2}) we are left with a fermion boundary term
proportional to
\be
(\d \q^1 \bar{\g}^{-} \q^1 - \d \q^2 \bar{\g}^{-} \q^2) {|},
\ee
which vanishes provided that $\W$ satisfies 
\be \label{cond}
\W^{t} \bar{\g}^{-} \W = \bar{\g}^{-},
\ee
where $\W^{t}$ is the transpose of $\W$. The fact that $\q^1$ and 
$\q^2$ have the same chirality implies that $\W$ should contain 
an even number of $\g$ matrices, and using the lightcone condition 
$\g^+ \q^{1,2}=0$ one finds that $\W$ may not contain 
$\g^+$ or $\g^-$. It then follows from (\ref{cond}) that $\W$ should be 
an orthogonal matrix, $\W^t \W=1$. 
Furthermore, the study of supersymmetry in later 
sections divides the discussion into 
$\W$ that satisfy one of the following two conditions
\be
D_-: \quad \W \Pi \W \Pi = -1, 
\qquad D_+: \quad \W \Pi \W \Pi = 1.
\ee
We will call the branes that satisfy the former  
condition ``$D_-$ branes'',
and the ones that satisfy the latter condition ``$D_+$ branes''.
As we will discuss in the next section, for a 
single D-brane the appropriate $\W$ 
is the product of gamma matrices $\g^{r'}$ where $r'$ is
transverse to the D-brane {\footnote{Other choices of $\W$  
are compatible with preserving closed string kinematical 
supersymmetries but not with dynamical supersymmetries. They will
also generically break worldvolume covariance. We will
not pursue such alternative boundary conditions here.}}. 

After fixing bosonic lightcone gauge ($x^+ = p^{+} \t$) and conformal
gauge ($ -g_{00} = g_{11} = 1$) the action becomes
\bea \label{laction}
S &=& T \int d^2\s \left ( p^{+} \del_{\t} x^{-} + \half (
(\del_{\t} x^{I})^2 - (\del_{\s} x^{I})^2 - m^2 (x^{I})^2 ) \right . \\
&& \left . + i p^+ (\q^1 \bar{\g}^{-} \del_{+} \q^1 + \q^2 \bar{\g}^{-} 
\del_{-} \q^2 - 2 m \q^1 \bar{\g}^{-} \Pi \q^2) \right ). \nn
\eea
Here $\del_{\pm} = \del_{\t} \pm \del_{\s}$ and $m = \mu p^{+}$. 
The field equations
for $(x^{I},\q^1,\q^2)$ in bosonic lightcone gauge and conformal gauge
($-g_{00} = g_{11}  =1$) are then 
\bea
(\del_{-} \del_{+} + m^2) x^{I} &=& 0; \nn \\
(\del_{+} \q^{1} - m \Pi \q^{2}) &=& 0; \label{fe} \\
(\del_{-} \q^{2} + m \Pi \q^{1}) &=& 0, \nn
\eea
to be solved together with the boundary conditions (\ref{bc1}) 
and (\ref{bc2}). 

The Virasoro constraints in bosonic lightcone gauge and in conformal
gauge for the metric are
\bea
T_{\s \t} &=& 
p^{+} [\del_{\s} x^{-}  + i  
( \q^1 \bar{\g}^{-} \del_{\s} \q^1 + {\q}^2 \bar{\g}^{-} 
\del_{\s} {\q}^2 )] + \del_{\t} x^{I} \del_{\s} x^{I}= 0; \label{vir}\\
T_{\t \t}&=& 
p^{+} [\del_{\t} x^{-} + i ({\q}^1 \bar{\g}^{-} \del_{\t} \q^1 
+ \q^2 \bar{\g}^{-} \del_{\t} {\q}^2 - 2 m \q^1 \bar{\g}^{-} \Pi \q^2)] \nn \\
&&- \half ( m^2 (x^{I})^2 - (\del_{\t} x^I)^2 
- (\del_{\s} x^{I})^2 ) = 0. 
\nn
\eea
From (\ref{v3}), once we fix both bosonic lightcone gauge and
conformal gauge, we get
\bea  \label{x+}
\d S &=& T \int d^2 \s \d x^{+} (p^+)^{-1} \left ( -\del_{\t} T_{\t \t}
+ \del_{\s} T_{\s \t}  + \del_{\t} x^{I} (\del_{+} \del_{-} x^{I} + m^2 x^{I})
\right . \\
&& \left . 
- 2 i p^+ [\del_{\t} \q^1 \bar{\g}^{-} 
(\del_{+} \q^1 - m \Pi \q^2) 
+\del_{\t} \q^2 \bar{\g}^{-} (\del_{-} \q^2 + m \Pi \q^1)] 
 \right ) 
\nn \\
&& - T \int d\t \d x^{+} \left ( \del_{\s} x^{-} + i 
(\q^1 \bar{\g}^{-} \del_{\s} \q^1 + \q^2 \bar{\g}^{-} \del_{\s} \q^2) 
\right ). \nn
\eea
In the bulk term we have rewritten terms involving $x^{-}$ using
$(T_{\t \t},T_{\t \s})$ from (\ref{vir}) whilst in the boundary term we have
already dropped terms which vanish because of the boundary conditions
imposed on $(x^{I},x^{+},\q^1,\q^2)$. Now to make the $x^+$ 
variational problem well defined, the boundary term should 
vanish. This means that 
\be \label{x-bc}
\left ( \del_{\s} x^{-} + i 
(\q^1 \bar{\g}^{-} \del_{\s} \q^1 + \q^2 \bar{\g}^{-} \del_{\s} \q^2) 
\right ){|} = 0
\ee
at the boundary. This condition is automatically satisfied provided
the Virasoro constraint $T_{\s \t}=0$ holds on the boundary. 
For both $D_-$ and $D_+$ branes the fermion
bilinear in (\ref{x-bc}) vanishes at the boundary\footnote{
To show this notice that for $D_-$ and $D_+$ branes
the boundary condition 
(\ref{bc2}) together with the field equation imply
$\pa_\s \q^1{|} = -\W \pa_\s \q^2{|}$ and
$\pa_\s \q^1{|} = (- \W \pa_\s \q^2 + 2 m \Pi \q^2){|}$, respectively.
In the latter case the vanishing of the bilinear in (\ref{x-bc}) 
follows upon using the fact that $\W \Pi$ is a symmetric matrix,
for the $\W$ given in (\ref{WB}).}, 
so $x^-$ satisfies a Neumann boundary condition.

Each term in the bulk variation involves either a Virasoro
constraint or a field equation. Thus the bulk variation 
term vanishes automatically
for all $\d x^{+}$ provided that we impose the other field equations
and the Virasoro constraints. One can understand this result as follows. 
Recall that given a conserved current ${\cal G}^a$  
that generates symmetry transformations $\d \f^A$, where $\f^A$ 
are all the fields in the theory, its divergence is equal to
\be \la{divcur}
\pa_a {\cal G}^a = \d \f^A {\d S \over \d \f^A} 
\ee
where it is understood that the parameter of the transformation
has been removed from  $\d \f^A$.\footnote{
Given a Lagrangian $L$ that transforms under a symmetry as
$\d L = \pa^a k_a$, for some $k_a$ (so that the action is 
invariant), the Noether current is given by 
${\cal G}^a = \pa L/(\pa (\pa_a \f^A)) \d \f^A - k^a$. Taking the 
divergence of this expression one gets (\ref{divcur}).}
In addition, in the presence of boundaries the spatial components of the 
current should vanish at the boundary. The derivation of the latter
is elementary and will be presented in section \ref{opst}.

Now let us apply this result to the current $T_{\a \t}$ that generates
worldsheet time translations, 
\be
\pa^\a T_{\a \t} = \pa_\t x^+ {\d S \over \d x^+} 
+ \pa_\t x^I {\d S \over \d x^I} 
+ \pa_\t \q^1 {\d S \over \d \q^1} + \pa_\t \q^2 {\d S \over \d \q^2} 
+\pa_\t x^- {\d S \over \d x^-}. 
\ee
In the light-cone gauge $\pa_\t x^+=p^+$, so 
\be \label{x+2}
{\d S \over \d x^+}=(p^+)^{-1} (\pa^\a T_{\a \t}
-\pa_\t x^I {\d S \over \d x^I}
-\pa_\t \q^1 {\d S \over \d \q^1}
-\pa_\t \q^2 {\d S \over \d \q^2}
-\pa_\t x^- {\d S \over \d x^-}).
\ee
It follows that the $x^+$ field equation is implied by the Virasoro
constraints and the other field equations (notice that in our case
$\d S/\d x^-$ is identically equal to zero). One can readily check 
that (\ref{x+}) agrees with (\ref{x+2}).
In addition, $T_{\s \t}$ should vanish at the boundary in order 
for the current $T_{a \t}$ to lead to a conserved current, and 
this gives exactly (\ref{x-bc}).

\subsection{Symmetry-related D-branes} \label{eqbr1}

Let $G$ be the isometry group of the target spacetime.
An isometry of the target spacetime 
may or may not leave invariant a given set 
of boundary conditions for open strings. 
Let $H$ be the subgroup 
of $G$ that respects the boundary conditions.
These transformations descend to symmetries
of the open string action and will be discussed
in the next section. On the other hand, the action of the 
coset generators $G/H$ results in 
new D-branes (i.e. open strings with a different
boundary condition), which however may be considered
as equivalent to the original ones since the two 
D-branes are related by 
an isometric reparametrization of the target spacetime.
In this subsection we discuss the branes that are
in the same equivalence class as the static branes; we
refer to them as symmetry related branes. 

The full set of (infinitesimal) bosonic isometries of the 
pp-wave background is as follows \cite{blau0} (fermionic isometries
will be discussed in the next section). 
Translations generated by $P^{\pm}$ and $P^I$ are given by
\bea
P^{\pm} &:& \qquad \d x^{\mp} = \ep^{\mp}; \\
P^I &:& \qquad 
\d x^{-} = \mu \sin \mu x^{+} \ep^{I} x^{I}, \hsp \d x^{I} = \cos \mu 
x^{+} \ep^{I}. \label{z0}
\eea
The rotational currents $J^{+I}$, $J^{ij}$ and $J^{i'j'}$ (where
$x^{i}$ and $x^{i'}$ parametrise $SO(4)$ and $SO(4)'$ respectively)
are associated with the respective transformations
\bea
J^{+I} &:& \qquad \d x^{-} = - \cos \mu x^{+} \ep^{I} x^{I}, \hsp \d x^{I} =  
\mu^{-1} \sin \mu x^{+} \ep^{I}; \label{rot} \\
J^{ij} &:& \qquad 
\d x^{i} = - \ep^{ij} x^{j}, \hsp \d \q = - \qu \ep^{ij} \g^{ij} \q; 
\label{z1}\\
J^{i'j'} &:& \qquad 
\d x^{i'} = - \ep^{i'j'} x^{j'}, \hsp \d \q = - \qu \ep^{i'j'} 
\g^{i'j'} \q, 
\eea
where $\ep^{ij} = - \ep^{ji}$ and $\ep^{i'j'} = - \ep^{j'i'}$.

Let us consider a finite $P^{r'}$ translation, 
\bea
\hat{x}^{-} &=& x^{-} - \mu \sin (\mu x^+) x^{r'} a^{r'} + 
\half \mu \sin (\mu x^{+}) \cos (\mu x^{+}) (a^{r'})^2; \\
\hat{x}^{r'} &=& x^{r'} - \cos(\mu x^+) a^{r'}, \nn
\eea
where $a^{r'}$ is a finite global parameter, and
the rest of the coordinates are unchanged.
One easily checks that this is an isometry:
$g_{\hat{x}^{\mu} \hat{x}^{\nu}} = g_{x^{\mu} x^{\nu}}$. 

Now we consider ``standard'' static boundary conditions 
in this new coordinate system, namely
\be
\del_{\s} \hat{x}^{+}{|} = 0; \hsp \hat{x}^{r'}{|} = 0; \hsp 
\del_{\s} \hat{x}^{-}{|} = 0.
\ee
These boundary conditions correspond to the following conditions
in our original coordinate system:
\be \la{mbc}
\del_{\s} x^{+}{|} = 0; \hsp
\del_{\s} x^{-}{|} = 
\mu \sin(\mu x^+) a^{r'} \del_{\s} x^{r'}{|}; \hsp
x^{r'}{|} =  a^{r'} \cos (\mu x^{+}){|}. 
\ee
Note that in the modified boundary conditions (\ref{mbc})
the Dirichlet position $x^{r'}$ is not constant but is explicitly
time dependent. This results of course from the explicit time
($x^+$) dependence of the translation Killing vectors. In the flat
space limit ($\mu \rightarrow 0$), the Dirichlet position
is purely constant, since the explicit time dependence drops
out of the isometry. 
%
%
\bigskip

One can also act with the other Killing vectors of the background on
the standard boundary conditions to generate modified boundary
conditions. The action of isometries along the worldvolume 
is trivial in the sense that this simply maps each point on the worldvolume
to another point on the worldvolume, without affecting the transverse
positions. Thus the boundary conditions are not affected by the action of
these isometries. These isometries generate symmetries
of the D-brane, and are discussed in the next section.

The only isometries which will generate distinct embeddings
are $P^{r'}$, $J^{+r'}$ and $J^{rr'}$. The action
of $J^{+r'}$ will be very similar to that of $P^{r'}$ (effectively
exchanging sines and cosines in the expressions given above). 
Thus a brane which is shifted by a combination
of $P^{r'}$ and $J^{+r'}$ transformations will have boundary
conditions:
\be
x^{r'}{|} = (a^{r'} \cos(\mu x^+) + b^{r'} \sin(\mu x^+)){|}. 
\label{rbc}
\ee
Here both $a^{r'}$ and $b^{r'}$ are constant vectors
in the Dirichlet directions. When these vectors
are orthogonal, say $a^{r'} = a$ and $b^{s'} = b$, 
the string traces out a trajectory of an ellipse 
in the $(r's')$ plane, with angular velocity $\mu$ with respect
to $x^+$. Thus these boundary conditions are appropriate to
describe a rotating brane. In particular, they describe
the branes arising from Penrose limits of giant gravitons.

When either of $a$ or $b$ are zero, the brane oscillates
along a line. (\ref{rbc}) represents the most general symmetry related
brane boundary condition,
up to the action of the surviving symmetries, and is
the condition we will use in what follows. The surviving
symmetries can shift, for example, $x^{+} \rightarrow x^{+} + x_{0}^{+}$
which changes the phase of the motion. 

The rotational symmetries $J^{IJ}$ in the pp-wave background 
act exactly as in flat space. Under a finite $J^{ij}$ transformation,
\be
\hat{x}^{i} = x^{i} \cos a - x^{j} \sin a, \hsp
\hat{x}^{j} = x^{j} \cos a + x^{i} \sin a,
\ee
where $a$ is a finite global parameter.  
If both coordinates are originally Dirichlet
they will both remain Dirichlet after the rotation. Furthermore,
if the transverse coordinates are zero, the rotation will
act trivially. However, if $i =r$ (Neumann) and 
$j = r'$ (Dirichlet) the boundary conditions will
be mixed after the rotation: as we said above, the broken
$J^{rr'}$ symmetry does act non-trivially. Since the $J^{rr'}$
transformations are independent of $\mu$ the discussion
of such rotated branes mirrors that in flat space (see
for example \cite{Berkooz:1996km}) and
we will not consider them further here. 

One should contrast the above discussion with the case of branes
in flat space. In that case, $\partial_{x^{r'}}$ is Killing
vector and one can use this isometry to position a D-brane
at any constant transverse position. In the pp-wave background,
however, it is $P^{r'}$ rather than $\partial_{x^{r'}}$
that is the Killing vector. Thus, even though the pp-wave
geometry is homogeneous, and any two points are 
related by a symmetry transformation, this does {\it not}
mean that {\it static} branes at distinct constant 
values of $x^{r'}$ are related by symmetry. 
The symmetry transformation relates static boundary conditions
to time dependent ones.

\subsection{Time-dependent D-branes}

We should note here that one can
also consider more general time dependent boundary
conditions for the open strings.
Such boundary conditions are 
consistent open string boundary conditions in bosonic 
lightcone gauge, both in flat space and in the pp-wave background.
In fact, {\it any} transverse boundary conditions of the form
\be 
x^{r'}{|} = x^{r'} (x^+){|} \label{arb}
\ee
are consistent with the field equations (\ref{fe}) in light cone gauge.
To see this first notice that the choice of bosonic light cone gauge
allows arbitrary $\d x^{+}$ on the boundary. Now consider
what variations $\d x^{r'}$ preserve the boundary conditions (\ref{arb}): 
we need to enforce
\be
p^{+} \d x^{r'}{|} = (\del_{\t} x^{r'}) \d x^{+}{|}, \label{cbc}
\ee
on the boundary. For the fermionic boundary conditions let us assume
the same condition as for the static D-branes holds, namely
(\ref{bc2}) holds
on the boundary, with appropriate $\W$ that satisfies
$\W \Pi \W \Pi=\pm 1$. Now if
we substitute these boundary conditions into the action variation
given in (\ref{v1}), (\ref{v2}), (\ref{v3}), (\ref{v4}) and
(\ref{v5}) we are left with the following boundary term
\bea
\d S &=& - \int d \t \left ( \del_{\s} x^{I} \d x^{I} 
+ \d x^{+} \del_{\s} x^{-} 
\right );  \\
&=& - (p^+)^{-1} \int d \t \d x^{+} \left ( \del_{\s} x^{I} 
\del_{\t} x^I + p^{+} \del_{\s} x^- \right ) 
= - (p^{+})^{-1} \int d \t \d x^{+} T_{\t \s}, \nn
\eea
where in the second line we have used $\del_{\s} x^{r}  =0$
for Neumann directions and (\ref{cbc}) for the transverse ``Dirichlet'' 
directions. So the boundary term again vanishes, provided that we impose
the Virasoro constraint $T_{\t \s}= 0$ on the boundary. 
We are then left with the usual field
equations (\ref{fe}) along with the boundary conditions (\ref{arb}) 
and (\ref{bc2}). Thus time dependent transverse 
boundary conditions are consistent
with the usual field equations (even in flat space, since nowhere
did we require $\mu \neq 0$) regardless of the explicit $x^+$ 
dependence. Furthermore, such boundary conditions also solve
the DBI field equations. It was 
already known in flat space that even arbitrary $x^{r'}(x^+)$ is consistent
with preservation of one quarter supersymmetry (half the kinematical
supercharges), see for example \cite{AFSS},
but later we will show that in fact such branes
admit a full compliment of sixteen supercharges.  

\bigskip

In this section we have considered a number of generalized
Dirichlet boundary conditions for open strings. 
We should also mention that it would be interesting to consider branes carrying
fluxes. It was found in \cite{ST} and \cite{TT} that certain constant fluxes 
can displace with $D$-branes from the origin while preserving
the same 16 supercharges as the brane at the origin 
and certain $D_+$ $5$-branes with non-constant fluxes preserved one quarter of 
the supersymmetry \cite{ST}. We leave for future work the 
exploration of the related open strings coupling to flux condensates. 

\section{Canonical Quantization}

In this section we present the canonical quantization 
of open strings with the various boundary conditions 
just discussed. It is useful to start by recalling 
the solutions of the field equations (\ref{fe}) without 
imposing any restrictions due to boundary conditions and/or the 
reality conditions of the fields. Recall that the 
most general solutions of the equation
\be \label{ffe}
(\del_{\t}^2 - \del_{\s}^2 + m^2 ) \f  = 0, 
\ee
are
\be \label{fn}
\f_k(\t,\s) = e^{-i(w_k \t + k \s)}, \qquad 
\tilde{\f}_k(\t,\s) = e^{-i(w_k \t - k \s)},
\ee
where $\w_k= {\rm{sgn}}(k) \sqrt{k^2 + m^2}$,
$k$ is an arbitrary complex parameter, 
and ${\rm{sgn}}(k)=\sqrt{k^2}/k$.
Notice that the choices $k=0$ and $k=\pm i m$ are special,
because in these cases $\f_k$ depends only on one of the two 
worldsheet coordinates,
\be
\f_0=e^{-i m \t}, \qquad \f_{im}=e^{m \s}, \qquad \f_{-im}=e^{-m \s}.
\ee
Given a solution of (\ref{ffe}) one can construct a 
solution of the fermionic equations as
\be \label{fereq}
\q^1 = i d_k \f_k \Pi \e_k + \tilde{\f}_l \tilde{\e}_l, 
\qquad \q^2 =\f_k \e_k  - i d_l \tilde{\f}_l \Pi \tilde{\e}_l,
\ee
where $\e_k$ and $\e_l$ are (independent) constant 
spinors, $k$ and $l$ are general complex parameters 
and 
\be
d_k={1 \over m} (\w_k - k). 
\ee

\subsection{Mode expansions}

The mode expansions for the closed and open strings 
with various boundary conditions follow
from the results we just presented by imposing
appropriate boundary conditions. 

For closed strings the complete set of solutions of 
(\ref{ffe}) is given by (\ref{fn}) with 
$k$ integral. The mode expansions of the bosonic
and fermionic coordinates were given in 
\cite{mt} and we summarise these results for
later convenience in appendix \ref{mcl}.

The explicit solution for the bosonic coordinates for 
a string with $(p-1)$ Neumann boundary conditions 
and ``generalized'' Dirichlet boundary conditions 
on the remaining coordinates is given by
\bea
x^{r}(\s,\t) &=& x_{0}^r \cos (m\t) + m^{-1} p_{0}^{r} 
\sin (m\t) + i \sum_{n \neq 0} \w_{n}^{-1} \a^r_{n} e^{-i \w_n \t} 
\cos (n \s); \label{neu} \\
x^{r'}(\s,\t) &=& x_0^{r'}(\s,\t)
+ \sum_{n \neq 0} \w_{n}^{-1} \a_{n}^{r'} e^{-i \w_{n} \t} \sin (n\s),
\label{dir}
\eea
where the zero mode part, $x_0^{r'}(\s,\t)$, satisfies 
(\ref{ffe}) and its exact form depends on the boundary conditions 
under consideration. The two types of zero modes we will
consider are the following. 
For a static brane located at $x_{0}^{r'}$ it is 
\be
x_0^{r'}(\s,\t)=
{x_{0}^{r'} \over e^{m \pi} + 1} (e^{m\s} + e^{m(\pi -\s)})
, \label{dir2}
\ee
For the symmetry related branes, 
as discussed in section \ref{eqbr1}, the appropriate zero
modes can be read off from (\ref{rbc}).
Reality of the bosonic coordinates requires that the zero modes
are real and 
\be
\a_{n}^{r} = (\a_{-n}^{r})^{\dagger}, \qquad
\a_{n}^{r'} = (\a_{-n}^{r'})^{\dagger}.
\ee
It is convenient to discuss separately $D_-$ and $D_+$ branes. 
The $\W$ appropriate to a single static D-brane 
is a product of the gamma matrices with indices
along transverse directions:
\be
\W = \prod_{r'=p}^{8} \g^{r'}. \label{omega}
\ee
For the corresponding anti-brane, $\W \rightarrow - \W$. 

\subsubsection{$D_-$ branes}

For these there are the following possibilities:
\bea
D7 &:& (+,-,2,4) \hsp (+,-,4,2) \nn \\
D5 &:& (+,-,1,3) \hsp (+,-,3,1) \\
D3 &:& (+,-,0,2) \hsp (+,-,2,0). \nn
\eea
In these expressions
we divide the eight transverse coordinates into two groups of
four, reflecting the $SO(4) \times SO(4)'$ symmetry of the background.
Following the notation of \cite{ST}, 
$(m,n)$ implies that there are $m$ Neumann directions in the
first group and $n$ in the second. 
Solving the fermion equations of motion, the mode expansions are
\bea
\q^1 &=& \q_0 \cos (m\t) + \tilde{\q}_{0} \sin (m\t) 
+ \sum_{n \neq 0} c_n \left ( i d_{n}\Pi \q_n \phi_n
+ \td{\q}_{n} \td{\f}_{n} \right ) ; \label{fer} \\
\q^2 &=& \Pi \td{\q}_0 \cos (m\t) - \Pi {\q}_{0} \sin (m\t) 
+ \sum_{n \neq 0} c_n \left ( - i d_n \Pi \td{\q}_n \td{\f}_n
+ {\q}_{n} \f_{n} \right ), \nn
\eea
where $\f_{n}$ and $\td{\f}_{n}$ are given in (\ref{fn}), $n$ 
is an integer and 
\be
c_n = {1 \over \sqrt{1+d_n^2}}.
\ee
Notice that the zero mode part is also of the form (\ref{fereq}).
In particular, if $\ep_0=\q_0 + i \tilde{\q}_0$ then the zero modes 
of $\q_1$ and $\q_2$ are given by $\q^1={\rm Re} (\f_0 \ep_0)$
and $\q^2={\rm Re} (- i d_0 \f_0 \Pi \ep_0)$.

The boundary condition enforces
\be
\td{\q}_{0} = -\W \Pi \q_0; \hsp \td{\q}_{n} = \W \q_{n}.
\ee
Combined with the chirality and light cone gauge constraints on $\q$
this reduces the number of independent zero modes to eight.
Reality of the spinors requires that
\be
\q_{-n} = (\q_{n})^{\dagger}.
\ee
For canonical quantisation we introduce canonical momenta
\be
P^{I} = \dot{x}^{I}, \hsp {\cal{P}}^{{\cal{I}}a} = \q^{{\cal I} a},
\ee
where ${\cal I} = 1,2$. Constraints on the fermionic phase space
can be incorporated using Dirac brackets. 
The classical Poisson-Dirac brackets are given by \cite{Met}
\bea
&&[P^{I}(\s),x^{J}(\s')]_{P.B.} = \pi \d^{IJ} \d(\s,\s'); \label{com} \\
&&\{(\q^1)^{\a}(\s),(\q^1)^{\b}(\s') \}_{D.B.} = \frac{i}{4 p^{+}} 
\pi (\g^+)^{\a\b} \d(\s,\s'); \nn \\
&&\{(\q^2)^{\a}(\s),(\q^2)^{\b}(\s') \}_{D.B.} = \frac{i}{4 p^+} 
\pi (\g^+)^{\a\b} \d(\s,\s'); \nn \\
&&[x_0^-, (\q^1)^\a]_{D.B} = {\pi \over 2 p^+} (\q^1)^\a, \quad
[x_0^-, (\q^2)^\a]_{D.B} = {\pi \over 2 p^+} (\q^2)^\a, \nn
\eea
where $x^-_0$ is the zero mode of $x_0^-$ which satisfies
$[p^+, x_0^-]=\pi$. All remaining brackets are zero. 
We now substitute the mode expansions into these relations
to determine the Poisson-Dirac brackets for the oscillators;
this involves decomposing the delta functions in terms of functions
which are orthogonal over the domain $[0,\pi]$. 
We give details of the calculation of oscillator commutation
relations in appendix D. 

Canonical quantization proceeds by replacing $\{. , .\}$ in
the Poisson-Dirac brackets by 
$-i [. , .\}$.  The (anti)commutators are found to be
\bea
[{a}_{n}^{I},a_{l}^{J}] &=& {\rm{sgn(n)}} \d_{n+l} \d^{IJ}, \hsp
[\bar{a}_{0}^{r},a_{0}^{s}] = \d^{rs}, \label{mode} \\
\{ \q^{\a}_{0},\q^{\b}_{0} \} &=& \frac{1}{8 p^+} (\g^{+})^{\a\b} \hsp
\{ \q^{\a}_{n}, \q_{m}^{\b} \} = \frac{1}{8 p^+}  
(\g^{+})^{\a\b} \d_{n+m,0}, \nn
\eea
with
\be
a_{0}^{r} = \frac{1}{\sqrt{2m}} (p_{0}^{r} + i m x_{0}^{r}), \hsp
\bar{a}_{0}^{r} = \frac{1}{\sqrt{2m}} (p_{0}^{r} - i m x_{0}^{r}), \hsp
{a}_{n}^{I} = \sqrt{\frac{1}{\left | \w_{n} \right |}} \a^I_n.
\ee
Notice that the coefficient of the fermionic anti-commutators
differs by a factor of two from the closed string result
(\ref{clcm}). This is a factor which is well-known in the flat space
limit. For the two types of generalized Dirichlet 
boundary conditions we consider, (\ref{dir2}) and 
(\ref{rbc}), the zero modes in the Dirichlet bosonic coordinates
are found to commute with themselves
and all of the other oscillators. Thus $x_{0}^{r'}$ and the
radii for the symmetry related branes, $a^{r'}$ and $b^{r'}$, 
are effectively c-numbers.

\subsubsection{$D_+$ branes}

In this case we have the following possibilities:
\bea \label{WB}
D9 &:& (+, -, 4,4) \nn \\
D7 &:& (+, -,3,3) \nn \\
D5 &:& (+, -,0,4) \hsp (+, -,2,2) \hsp (+, -, 4,0) \\
D3 &:& (+, -, 1,1) \nn \\
D1 &:& (+, -, 0,0). \nn
\eea
We should add the caveat here that $(+, -,0,4)$ and $(+, -, 4,0)$ constant
embeddings do not solve the DBI field equations without the 
introduction of worldvolume flux \cite{ST}. Nevertheless, these
branes appear to be on equal footing with the other ones as 
far as the analysis in this paper is concerned.

For branes in (\ref{WB}) the mode expansion of the bosons is exactly 
as for the $D_-$ branes and for the fermions we find that\footnote{
Note that possible terms in the mode expansion of the form
$\q_{k} e^{i \w_k \t \pm k \s}, $
with $k^2 \neq m^2$ are excluded by the boundary conditions.}  
\bea
\q^1 &=& \q_0^{+} e^{m\s} + {\q}_{0}^{-} e^{-m\s} + 
\sum_{n \neq 0} c_n \left ( i d_n \Pi \q_n \phi_n
+ \td{\q}_{n} \td{\f}_{n} \right ) ; \label{fer2} \\ 
\q^2 &=& \Pi \q_0^{+} e^{m \s} - \Pi {\q}_{0}^{-} e^{-m\s}
+ \sum_{n \neq 0} c_n \left ( - i d_n \Pi \td{\q}_n \td{\f}_n
+ {\q}_{n} \f_{n} \right ), \nn
\eea
Notice that the zero mode part can be rewritten in the same 
form as the non-zero modes, but now instead of $\f_0$ one should 
use $\f_{im}$ and $\f_{-im}$ (using the notation introduced in
(\ref{fn})).

The boundary condition enforces
\bea
\q_{0}^{\pm} &=& \pm \W \Pi \q_{0}^{\pm}; \label{zero1} \\ 
\td{\q}_{n} &=& c_{n}^2 (\W (1 - d_{n}^2) - 2 i d_{n} \Pi) \q_{n}. \nn
\eea
Reality of the fermions is enforced by 
\be
\q_{-n} = (\q_{n})^{\dagger},
\ee
with $\q_{0}^{\pm}$ real.

Canonical quantization proceeds as before by introducing canonical 
momenta; details of the analysis are given in the appendix. 
The (anti)commutators are given by
\bea
[{a}^{I}_{n}, a_{l}^J] &=& {\rm{sgn}}(n) \d_{n+l,0} \d^{IJ}, \hsp 
[\bar{a}_{0}^{r},a_{0}^s] = \d^{rs}; \nn 
\\
\{ \q^{\a}_{n}, \q^{\b}_{m} \} &=& \frac{1}{8 p^+} (\g^{+})^{\a\b}
\d_{n+m,0}; \\ 
\{ \caP_{+} \q^{+}_{0}, \caP_{+} \q^{+}_{0} \} &=&  
\frac{\pi m}{4 p^+ (e^{2\pi m}  - 1)} \caP_{+} \g^{+}; \nn \\
\{ \caP_{-} \q^{-}_{0}, \caP_{-} \q^{-}_{0} \}  &=&   
\frac{\pi m e^{2 \pi m}}{4 p^+ (e^{2 \pi m} - 1)}
\caP_{-} \g^{+}. \nn 
\eea
We include in these brackets the projections
\be 
\caP_{\pm} = \half ( 1 \pm \W \Pi),
\ee
which enforce the constraints on the zero modes. Note that in the
$m \rightarrow 0$ limit, the fermion zero mode terms reproduce
those in the previous section, as they must.
Again the zero modes in the Dirichlet coordinates do not
have conjugate momenta and commute with all of the oscillators.

\section{Worldsheet symmetries and conserved charges}

In this section we discuss the construction of symmetry generating
charges from conserved currents in the presence of worldsheet boundaries.
In a flat target space, one often obtains these using a 
proper doubling of the interval so that classical open string
solutions satisfy periodic boundary conditions for $[0,2\pi]$; one
then substitutes the mode expansions into the expressions for
the closed string supercharges. Here it is not obvious 
how to apply such a doubling trick since the mode
expansions involve non-periodic functions (see, for instance, 
(\ref{fer2})).
Instead, we will consider the symmetries of the worldsheet 
action with given boundary conditions from first principles. 

In sigma models the isometries of the target spacetime
result in conserved worldsheet currents. However, these
are not the only global symmetries. In our case the 
light-cone action is quadratic in the fields and
this implies that there are an infinite number of global 
worldsheet symmetries. Particular cases of these
symmetries are some of the target space kinematical 
symmetries (to be discussed below). 

If the closed string action is invariant under a given set of symmetry
transformations, the open string action will only be invariant
under a subset of these transformations which is compatible with the
(given) boundary conditions on the worldsheet fields. A novel 
feature for our open strings is that in some cases it is a combination
of target space and worldsheet symmetries that is 
compatible with the chosen boundary conditions.

\subsection{Closed strings} \label{cst}

To obtain the symmetry currents we use the standard Noether method based
on localisation of the parameters of global transformations. It is useful to
first recall the symmetries of the action (\ref{laction}) when the string 
worldsheet is closed. Suppose that
$\ep$ is a parameter of a global transformation that leaves the action 
invariant (up to total derivative terms which vanish when the
worldsheet is closed); then when $\ep$ becomes local the variation 
of the action takes the form 
\be
\d S = T \int d^2 \s {\cal G}^{a} \del_{a} \ep, \label{var1}
\ee
where ${\cal G}^a$ is the corresponding symmetry current.
As usual this implies that there is a conserved symmetry current,
$\del_{a} {\cal G}^a = 0$, with a corresponding
conserved charge given by 
\be
G = T \int^{2 \pi}_{0} d \s {\cal G}^{\t}.
\label{var2}
\ee
We take all generators to be Hermitian.
We normalize the currents such that the Poisson bracket of a bosonic
generator with a field $\Phi$ yields 
the corresponding symmetry variation, $\d \Phi$,
but the Poisson bracket of a fermionic generator yields, $-i \d \Phi$.
The factor of $i$ is required by the reality conditions; alternatively 
one can consider anti-Hermitian fermionic generators.

The isometries of the target space give rise to global worldsheet symmetries.
The derivation of the transformation rules and of the corresponding
Noether currents and charges has been given by Metsaev in
\cite{Met}. What we would like to point out here is that 
some of these symmetries admit an infinite extension.

Let us consider the transformation
\be \label{afftr}
\d x^I = \ep^{I} \f(\t,\s), \label{sm1} \qquad
p^+ \d x^-= - x^I \ep^I \pa_\t \f(\t,\s).
\ee
The Lagrangian in (\ref{laction}) is invariant (up to spatial 
total derivatives) under this symmetry provided
$\f(\t,\s)$ satisfies
\be \label{ffe1}
(\del_{\t}^2 - \del_{\s}^2 + m^2 ) \f  = 0. 
\ee
The associated conserved currents are 
\bea
P_{\f}^{I \t} &=& \del_{\t} x^{I} \f - x^{I} \del_{\t} \f;  \label{pcur}\\
P_{\f}^{I \s} &=& x^{I} \del_{\s} \f - \del_{\s} x^{I} \f. \nn 
\eea
For the closed string, only those $\f$ which respect the periodicity of
the worldsheet give rise to conserved charges. The complete 
set of solutions of (\ref{ffe1}) subject to periodic boundary conditions
are given by $\f_n$ and $\td{\f}_{n}$, introduced in (\ref{fn}),
with $n$ an integer. 
It follows that there is a (countably) infinite number of symmetries.
For each $n$ there are two (complex) conserved charges, the first being
\be \label{Pn}
P^{1 I}_{n} = T \int d \s (\del_{\t} x^{I} \td{\f}_{-n} 
- x^{I} \del_{\t} \td{\f}_{-n}), \label{pn}
\ee
and the other $P^{2 I}_n$, which is the same as $P^{1 I}_n$ but
with $\td{\f}_{-n}$ replaced by ${\f}_{-n}$. Notice that 
complex conjugation relates $P^{{\cal I} I}_{n},{\cal I}=1,2 $ to 
$P^{{\cal I} I}_{-n}$. 

The $n=0$ case is special in that we recover 
symmetries associated with the isometries of the target space.
Using $\f=\cos m \t$, the transformation (\ref{afftr})
reduces to the translation isometry given in (\ref{z0}), and 
using $\f=\m^{-1} \sin m \t$ 
we obtain the rotational isometry given in (\ref{rot}). 
In terms of currents
\be
P^I = {\rm Re} P_0^I, \qquad J^{+I} = {\rm Im} P_0^I.
\ee
Notice that only in the $n=0$ case does the function $f$ 
depend only on worldsheet time, and so by using the lightcone
relation, $x^+ = p^+ \t$, the symmetries can be expressed
in terms of target space coordinates. All the other symmetries
are stringy extensions. 

There is a corresponding story involving the fermions:
a shift in the fermions by a parameter that satisfies the
fermionic field equation is a symmetry of the action (\ref{laction}).
There is again a countably infinite number of such transformations,
\bea \label{Q+n}
\d \q^{1} & = & - c_{n} (i d_n \Pi \ep^2_n  \f_n  + \ep^1_n \tilde{\f}_n); \\
\d \q^{2} & = & - c_{n} (- i \Pi d_{n} \ep^1_{n} \tilde{\f}_{n}
+ \ep^2_n  \f_n); \nn \\
\d x^- &=& i (\q^1 \bar{\gamma}^- \d \q^1 
+ \q^2 \bar{\gamma}^- \d \q^2), \nn
\eea
where $m d_{n} = (\w_{n} - n)$, $c_n=(1+d_n^2)^{-1/2}$,
the constant spinors $\e_n$ satisfy $\bar{\g}^+ \ep_n=0$, 
$n>0$ is integral ($n=0$ is considered below).
The current associated with the  $\ep_n^1$ symmetry is
\bea \label{Qn1}
Q_{-n}^{1 \t} &=& 2 p^+ \bar{\g}^{-} (\q^{1}
- i d_{n} \Pi \q^{2}) c_n \tilde{\f}_{n}; \\
Q_{-n}^{1 \s} &=& 2 p^+  \bar{\g}^{-} (\q^{1} 
+ i d_{n} \Pi \q^{2}) c_n \tilde{\f}_{n}, \nn
\eea
and that associated  with $\ep_n^2$ symmetry is
\bea \label{Qn2}
Q_{-n}^{2 \t} &=& 2 p^+ \bar{\g}^{-} (\q^{2}
+ i d_{n} \Pi \q^{1}) c_n \f_{n}; \\
Q_{-n}^{2 \s} &=& - 2 p^+ \bar{\g}^{-} (\q^{2} 
- i d_{n} \Pi \q^{1}) c_n \f_{n}. \nn
\eea
Both currents are conserved when the fermions are on-shell. 

The $n=0$ case is special. In this case
the transformations are given by
\bea
\d \q^1 &=& \cos m \t \ep^2 - \sin m \t \Pi \ep^1; \nn \\
\d \q^2 &=& - (\cos m \t \ep^1 + \sin m \t 
\Pi \ep^2); \label{z2} \\
\d x^{-} &=& i (\q^1 \bar{\gamma}^- \d \q^1 
+ \q^2 \bar{\gamma}^- \d \q^2). \nn
\eea
Since these only depend on $\t$ they can be written entirely
in terms of target space coordinates. These transformations
are precisely the kinematical supersymmetry transformations
obtained in \cite{Met}. The corresponding Noether currents are 
\bea
Q^{+1 \t}&=&2 p^+ \bar{\gamma}^- (\cos \m x^+  \q^2 
+ \sin \m x^+ \Pi \q^1); \nn \\
Q^{+1 \s}&=&2 p^+  \bar{\gamma}^- (-\cos \m x^+ \q^2 
+ \sin \m x^+ \Pi \q^1); \\
Q^{+2 \t}&=&-2 p^+ \bar{\gamma}^- (\cos \m x^+ \q^1 
- \sin \m x^+ \Pi \q^2); \nn \\
Q^{+2 \s}&=&-2 p^+  \bar{\gamma}^- (\cos \m x^+ \q^1 
+ \sin \m x^+ \Pi \q^2); \nn
\eea
These expressions agree with the ones given in \cite{Met}.

The action (\ref{laction}) is also invariant under 
the (dynamical) supersymmetry transformations
\bea
\d x^{I} &=&  
2 i (\q^1 \bar{\g}^{I} \ep^1 + \q^2 \bar{\g}^{I} \ep^2); \label{z3} \\
p^+ \d \q^1 &=& \half ( \g^{+ I} \del_{-} x^{I} \ep^1 + 
 m \Pi \g^{+I} x^{I} \ep^2); \nn \\
p^+ \d \q^2 &=& \half ( \g^{+ I} \del_{+} x^{I} \ep^2 - 
m \Pi \g^{+I} x^{I} \ep^1), \nn 
\eea
which were also given in \cite{BPZ}. 
Just as in the flat space transformations ($m = 0$), these 
transformations involve spinors $\ep$ which do not satisfy the fermionic
lightcone gauge condition $\bar{\g}^{+} 
\ep = 0$.  The $\ep$ transformations effectively have to be
combined with a kappa symmetry transformation in order to
preserve $\bar{\g}^{+} \q = 0$; this is the origin of the $\g^{+}$ 
factors in the transformations and of the simultaneous $x^I$ 
transformations. 
The corresponding currents are given by 
\bea
Q^{-1 \t} &=& 2 (\del_{-} x^{I} \bar{\g}^{I} \q^1 
- m x^{I} \bar{\g}^{I} \Pi \q^2 ); \nn \\
Q^{-1 \s} &=& 2 (\del_{-} x^{I} \bar{\g}^{I} \q^1 
+ m x^{I} \bar{\g}^{I} \Pi \q^2 ); \\
Q^{-2 \t} &=& 2 
(\del_{+} x^{I} \bar{\g}^{I} \q^2 
+ m x^{I} \bar{\g}^{I} \Pi \q^1 ); \nn \\
Q^{-2 \s} &=& 2 
(- \del_{+} x^{I} \bar{\g}^{I} \q^2 
+ m x^{I} \bar{\g}^{I} \Pi \q^1 ). \nn 
\eea
These expressions agree with the ones given in \cite{Met}.

The closed string (super)charges generate the superalgebra of the 
plane wave background, which is given in appendix B. 
The realization of this algebra in terms of closed string modes
was worked out in \cite{mt}. As we have seen the
kinematical charges, $P^I, J^{+I}$  and  $Q^{+1}, Q^{+2}$, 
are members of an infinite family of symmetries and
thus one may extend the superalgebra to include these
charges as well. The charges $P_m^{{\cal I} I}, Q_n^{{\cal I} I}, 
{\cal I}=1,2$, when evaluated on-shell 
using the closed string mode expansions reviewed in
appendix C are given by
\bea
&&P^{1I}_{n} = 2 \a^{1 I}_{n}, \qquad P^{2 I}_{n} =2  \a^{2 I}_{n}, \nn \\
&&Q_n^1 = 2 p^+ \bar{\g}^- \q_n^1, \qquad
Q_n^2 = 2 p^+ \bar{\g}^- \q_n^2. 
\eea
Using these one can compute the extension of the superalgebra.
This result will be given in \cite{ST3}.

We should comment that although we have identified all conserved charges 
realised linearly in the oscillators there are also additional 
conserved charges realised non-linearly in the oscillators. 
As can be seen from formulas in appendix \ref{mcl}, some 
of the target space symmetry generators are realized in terms of 
oscillator bilinears. There are, however, additional symmetries
whose on-shell charge is bilinear in the oscillators.
An example is the symmetry
\bea \la{fr}
\d \q^1 &=& \e^{IJKL} \g^{IJKL} \pa_{\s} \q^1; 
\\
\d \q^2 &=& \eta \e^{IJKL} \g^{IJKL} \pa_{\s} \q^2, \nn
\eea
where $\e^{IJKL}$ is the parameter of the transformation antisymmetric
in all indices (note that there is no summation over 
$I,J,K,L$ in (\ref{fr})).
$\eta$ is equal to plus one for $[\Pi, \g^{-IJKL}] = 0$ and minus
one for $\{ \Pi, \g^{-IJKL} \} = 0$. The associated tensorial charge is 
\be \la{ten} 
J^{IJKL} \sim \int d\s (\pa_{\s} \q^1 \g^{-IJKL} \q^1
                              + \eta \pa_{\s} \q^2 \g^{-IJKL} \q^2),
\ee
and on-shell is equal to 
\be \la{thre}
J^{IJKL} \sim \sum_{n} n (\q^{1}_{-n} \g^{-IJKL} \q^1_n - \eta \q^2_{-n} 
\g^{-IJKL} \q^2_{n}).
\ee
There are also other closely related tensorial charges. We will 
discuss all of them as well as the corresponding extension
of the closed string superalgebra together with the 
related open string charges in \cite{ST3}.

\subsection{Open strings} \label{opst}

Now we consider symmetries of the open string action,
with given boundary conditions on the worldsheet fields. Only
those symmetries of the closed string action which are compatible 
with these boundary conditions will leave the open string action 
invariant. Suppose we consider an infinitesimal global symmetry
transformation of the form
\be
\d z^{m} = H^{m}(z^n) \ep.
\ee
Only when this preserves the boundary conditions on all the fields
$z^{m}$ does this leave the action (including boundary terms) 
invariant. To take a specific example, suppose $x^{r'}$ satisfies
a Dirichlet boundary condition, $\del_{\t} x^{r'}{|} = 0$. Then 
a global $P^{r'}$ transformation manifestly does not preserve
this condition. In fact, if we
evaluate the variation of the action under such a global transformation 
(using (\ref{v1}) and so on), we find that there is a non-vanishing
boundary term
\be
\d S = - \frac{1}{\pi} \int d \t \del_{\s} x^{r'} 
\cos(\mu x^{+}) \ep^{r'},
\ee
explicitly demonstrating that the action is not preserved. 

Symmetries of the open string action will thus only be those which
preserve the boundary conditions and for which, by construction,
the variation of the action under a local transformation is
\be
\d S = \frac{1}{\pi} \int d^2\s  {\cal G}^{a} \del_{a} \ep,
\ee
where the ${\cal G}^a$ corresponding to each preserved symmetry
of the action is of the same form as for the closed string.
Partially integrating this expression, we get
\be
\d S = - \frac{1}{\pi} \int d^2 \s \del_{a} G^{a} \ep 
+ \frac{1}{\pi} \int d \t \left [ G^{\s} \ep \right ]^{\s = \pi}_{\s = 0}.
\ee
On shell the bulk term vanishes by virtue of the equations of motion.
If this is to be a good symmetry given the boundary conditions the
boundary term must also vanish for all $\ep$. This requires
that 
\be
[G^{\s}]_{\s = \pi} = [G^{\s}]_{\s = 0} = 0, \label{expl}
\ee
which will hold by construction if the variational problem has been
set up correctly (so all boundary terms vanish) and if 
the conserved symmetries have been correctly identified.
Thus it is a consistency check
once we have identified the preserved symmetries for specific
boundary conditions that this term does indeed vanish. The
conserved charges are then given by
\be
G = \frac{1}{\pi} \int d \s G^{\t}.
\ee
These are conserved since
\be
\pi \del_{\t} G = \int d \s \del_{\t} G^{\t} 
= - \int d \s \del_{\s} G^{\s} = - [G^{\s}]^{\s = \pi}_{\s = 0} = 0,
\ee
where the last equality holds by construction. 
Let us now discuss in detail which are the preserved charges
for each of the boundary conditions given in the previous section.

\subsubsection{D-branes}

In order to respect the boundary conditions chosen 
for the D-brane the symmetry transformations 
should be such that $\d x^{r'}{|} = 0$ and
$\d \q^1{|} = \W \d \q^2{|}$ at the boundary; 
the other variations $\d x^{+}{|}$, $\d x^{-}{|}$ and 
$\d x^r{|}$ 
are arbitrary. Looking at the symmetry transformations of the closed
string action in (\ref{z0}), (\ref{z1}), (\ref{z2}) and (\ref{z3}), it is 
already apparent that some of these variations are not compatible
with these conditions. As we have discussed, under a global transformation 
corresponding to $P^{r'}$, the transformation
does not preserve $\d x^{r'}{|} = 0$ on the boundary, and hence
does not preserve the action (including boundary terms).
Thus $P^{r'}$ cannot be a conserved charge of the D-brane.
This is of course apparent since the D-brane explicitly breaks
the translational symmetry.

The other translational charges $P^{+}$, $P^{-}$ and $P^r$ are
compatible with the boundary conditions and are conserved
charges of the D-brane. As mentioned above it is a nice consistency
check to explicitly verify that (\ref{expl}) holds for these charges.
Take for example $P^r$: then using the explicit forms for
the currents given in the appendix
\be
{\cal P}^{r \s} = - (\cos (\mu x^{+}) \del_{\s} x^{r} + \mu
\sin (\mu x^+) x^{r} \del_{\s} x^+).
\ee
This vanishes on the boundary by virtue of the Neumann conditions
on both $x^+$ and $x^r$. In what follows we have explicitly checked
that each charge which we identify as being conserved satisfies $G^{\s}(0)
= G^{\s}(\pi)=0$. 

Similarly to $P^I$, some of the rotational $J^{+I}$ symmetries 
are preserved by the boundary conditions, i.e. the $J^{+r}$ ones, and some
are broken, i.e. the $J^{+r'}$ ones. In the closed string case
we showed that $P^I$ and $J^{+I}$ were the $n=0$ members of an infinite
number of symmetries. The same holds for open strings
as well. The argument for the symmetries is the same as that
given in the closed string case, i.e. a shift in the coordinates 
$x^I$ by a parameter $\e^I$ that satisfies the $x^I$ field equations
is a symmetry of the worldsheet action provided the boundary
terms vanish. Consider first the case of transformations
that involve a shift of the Neumann coordinates,
\be \label{opN} 
\d x^r = \e^r f_n,  \qquad p^+ \d x^-= - x^r \ep^r \pa_\t f_n 
\ee
where $\sqrt{ \left | \w_n \right | } f_n = e^{- i \w_n \t} \cos n \s$
(the prefactors are introduced for later convenience).
This transformation leaves the action invariant including 
boundary terms.
The corresponding conserved charge is given by (\ref{Pn}) 
but with $\td{\f}_{-n}$ replaced by $f_n$. 

Consider now the similar transformation involving the 
Dirichlet direction, i.e. (\ref{opN}) but with $x^r$ replaced
by $x^{r'}$ and $f_n$ replaced by
$ \sqrt{ \left | \w_n \right |}
\tilde{f}_n= - ie^{- i \w_n \t} \sin n \s$. This transformation,
however, leads to a non-zero boundary term if 
$x_0^{r'}(\s,\t)$ is non-zero at the string endpoints,
where $x_0^{r'}(\s,\t)$ is defined in (\ref{dir}).
So unless the brane is located at the origin of transverse space,
these transformations do not lead to a conserved charge.
However, one can modify the transformation rules
such that there is conserved charge even when the 
brane is located at arbitrary constant position.
Indeed, consider the transformations
\be \label{opD} 
\d x^{r'} = \e^{r'} \tilde{f}_n,  \qquad 
p^+ \d x^-= -  \ep^{r'} (x^{r'} \pa_\t \tilde{f}_n 
+ P_{(0) \tilde{f}_n}^{r' \t})
\ee
where $P_{(0) \tilde{f}_n}^{r' \t}$ is given by (\ref{pcur}) 
but with $x^{r'}$ replaced by $x_0^{r'}(\s,\t)$. 
One may verify that these transformations leave the 
action invariant, including boundary terms, even when 
the brane is located at a non-zero transverse position.
The corresponding conserved charge is then given by (\ref{pn})
but with $x^{r'}$ replaced by $(x^{r'} - x_0^{r'})$ and
$\td{\f}_{-n}$ by $\tilde{f}_n$. It follows that when the 
charge is evaluated in terms of modes it takes the same form
as for the conserved charge associated with the brane located at 
the origin of transverse space. 

This is a phenomenon that we will see repeatedly in what follows: certain
symmetries appear to be broken when the brane is located 
away from the origin of transverse space, and the breaking terms
solely depend on quantities fully determined by the boundary conditions
(in the case above the breaking terms depend only on the 
boundary value of the Dirichlet coordinates). In these cases, however,
one can find modified transformation rules (involving worldsheet
symmetries) that lead to a conserved charge. 
The latter when evaluated in terms of modes
is exactly equal to the conserved charge of the brane located 
at the origin.

The $n=0$ case is special, as for the closed string case. 
In this case the real and imaginary part 
of the transformations in the Neumann direction give the $P^r$ and
$J^{+r}$ transformations. In addition, the Dirichlet directions
do not yield any conserved charge when $n=0$, 
in agreement with the fact that the D-branes break the $P^{r'}$ 
and $J^{+r'}$ symmetries.

\bigskip
  
We now move to the rotational symmetries $J^{IJ}$. Consider
first the $J^{rs}$ case. The
fermionic variations are compatible with the boundary conditions
since
\be
\d \q^1{|} = -\qu \e^{rs} \g^{rs} \q^1{|} 
= - \qu \ep^{rs} \g^{rs} \W \q^2{|} = \W \d \q^2{|}, \label{fv}
\ee
where in the second equality we use the relation between
fermions on the boundary and in the final equality we use 
$[\W, \g^{rs}]=0$. This follows from the explicit form
of $\W$ given in (\ref{omega}).
Since $\d x^{r}$ is arbitrary
on the boundary the bosonic variations are also compatible
with the boundary conditions and $J^{rs}$ is a conserved charge.
However, the $J^{rr'}$ symmetry is broken since the bosonic variations
mix Dirichlet and Neumann coordinates (and furthermore
the fermionic variations do not preserve the fermion boundary
conditions for reasons akin to above). 

The $J^{r's'}$ symmetry is another instance where we
will use the worldsheet symmetries we just discussed 
in order to restore a seemingly
broken symmetry. The fermionic variations do preserve the fermion boundary
conditions (as in (\ref{fv}), $\W$ commutes with $\g^{r's'}$)
and there is actually a conserved current associated
with these transformations. 
The bosonic variations, however, will only preserve the boundary condition
if the brane is located at the origin. That is, $\d x^{r'}{|} = 0 = 
\ep^{r's'} x^{s'}{|}$ can be satisfied on the boundary only if
$x^{s'}_{0} = 0$. So even though there is a conserved current for
any value of transverse position, $x^{r'}_{0}$, the transformation
rules change that value. Thus, the $J^{r's'}$ symmetry 
is broken by the position of the brane unless the brane 
is located at the origin of transverse space. 

We will now see that even at arbitrary transverse position
one can obtain a good symmetry by combining the $J^{r's'}$
symmetry with a certain worldsheet symmetry.
Consider the worldsheet symmetry given by 
$\d x^{r'} = \e^{r' s'} x_0^{s'}(\s,\t)$,
where $x_0^{r'}(\s,\t)$ is defined in (\ref{dir}).
By construction, $x_0^{s'}(\s,\t)$ satisfies (\ref{ffe}),
so this transformation leaves the action
invariant\footnote{In general, the fact that the parameter satisfies
the free field equation only implies that the Lagrangian 
is invariant up to total derivatives. The cancellation of the 
total derivative terms depends on the boundary conditions
imposed on the fields. The worldsheet symmetry $\d x^{r'}$
in actually a symmetry including boundary terms. Notice that 
the $J^{r's'}$ transformations also leave the action invariant
including boundary terms.}. We now combine the $J^{r's'}$ symmetry with 
this worldsheet symmetry by considering the transformations,
 \be
\d x^{r'} = \ep^{r's'} (x^{s'} - x_{0}^{s'}(\s,\t)); \hsp \d \q = - \qu 
\ep^{r's'} \g^{r's'} \q.
\ee
These transformations leave invariant the action, including
boundary terms, and respect both fermionic and bosonic 
boundary conditions. It follows that this is a good symmetry.
The associated conserved current is just $J^{r's'}$, with 
$x^{r'}$ replaced by $(x^{r'} - x_{0}^{r'}(\s,\t))$ and similarly
for $x^{s'}$. We shall show in \cite{ST3} that the conserved charge when 
realised in terms of the oscillators is precisely that for the 
branes at the origin. 

\bigskip

In the above we have used worldsheet symmetries
to subtract from symmetry currents terms which
do not preserve the boundary conditions. However this trick
can be applied much more generally: we may also combine 
good worldsheet symmetries with good target space symmetries
(i.e. target space symmetries that are compatible
with the boundary conditions). In this way one
can obtain Noether charges that are equal 
to the original charge associated with a superisometry
but with certain terms subtracted. 

A specific example that will arise in the open string 
algebras in \cite{ST3} is
the following. Suppose we first isolate the part of the 
(conserved) Hamiltonian symmetry
current which involves only the Dirichlet zero modes and which
is conserved by itself; call this
$\D H^{a}$. Then under a symmetry transformation
\be
\d x^{+} = - \ep^+; \hsp
p^{+} \d x^{-} = - \D H^{\t} \ep^{+},
\ee
the action is preserved and the corresponding Noether charge
is
\be
\hat{H} = H - \frac{1}{\pi} \int^{\pi}_{0} d \s \D H^{\t} \equiv
H - \D H, \la{shft}
\ee
which when evaluated in terms of modes is precisely the original
Hamiltonian with the Dirichlet zero mode (c-number) contribution subtracted.

\bigskip

We next consider the transformations associated with the supercharges.
As in previous sections, it will be convenient to discuss 
separately the $D_-$ and $D_+$ branes.

\subsubsection{$D_-$ branes}

First consider the kinematical supercharges $Q^+$.
The corresponding transformation rules are given in (\ref{z2}). 
Under global transformations corresponding to $Q^+$ the
action is invariant up to the total derivative term 
\be
\d S = \frac{1}{\pi} \int d \t 
i p^+ (\d \q^1 \bar{\g}^- \q^1 - \d \q^2 \bar{\g}^- \q^2).
\ee
Thus provided that the variations are compatible with the 
fermion boundary conditions, that is, $\d \q^1 = \W \d \q^2$
on the boundary, the action remains invariant.
We find that the fermion boundary conditions are compatible 
with (\ref{z2}) only when 
\be
\ep^2 = - \Omega \ep^1, \hsp \Pi \W \Pi \W = -1. \label{c1}
\ee
Note that the second condition would not have to be imposed
in flat space ($\mu = 0$) but is necessary here to enforce
$\d \q^1{|} = \W \d \q^2{|}$ at all $x^+$. Thus 
the boundary conditions
can preserve at most one half of the $Q^+$ supercharges, the
combinations 
\be
q^{+} = \half (-Q^{+2} + \bar{\W} Q^{+1}).
\ee
(The normalization of $q^+$ is for later convenience). We conclude that 
the $D_-$ branes preserve one half of the kinematical 
supersymmetries of the closed string. Let us note at the this
stage that these symmetries are preserved regardless of the
Dirichlet boundary conditions: even $D_-$ branes with general time
dependent boundary conditions $x^{r'}(x^+)$ preserve
these same charges.

The $Q^+$ symmetry preserved by the $D_-$ 
branes is just the $n=0$ member of 
a family of infinite number of symmetries. The additional 
symmetry transformations are 
given by (\ref{Q+n}) but with parameters satisfying
\be \label{opQ+}
\ep_n^1 = \W \ep_n^2.
\ee
The corresponding Noether charges can be obtained from (\ref{Qn1})
and (\ref{Qn2}) by taking combinations 
$Q_n = \half (Q_{n}^1 + \bar{\W} Q_{n}^2)$. 

\bigskip

Let us now consider the dynamical supercharge ($Q^{-}$) 
transformations. Once again the
action is only invariant up to a total derivative term and 
the boundary conditions are not in general consistent with (\ref{z3}). 
Let us consider first the $\d x^{I}$ transformations: if these
are to preserve the Dirichlet boundary conditions then we must
consider only such transformations satisfying
\be
\d x^{r'}{|} = 0 
= \q^2{|} (\W^{t} \bar{\g}^{r'} \ep^1 + \bar{\g}^{r'} \ep^2) \hsp 
\rightarrow \hsp \ep^1 = \W \ep^2, \label{rel1}
\ee
where we have used $\{ \W,\g^{r'} \} = 0$ (which follows from 
(\ref{omega})). Thus the
D-brane can preserve at most one half of the $Q^{-}$ supersymmetries.
Now under transformations (\ref{z3}) the action is invariant up to
the boundary term
\be
- \int d \t \left (\del_{+} x^{I} \q^1 \bar{\g}^{I} \ep^1 - \del_{-} 
x^{I} \q^2 \bar{\g}^{I} \ep^2 + m x^{I} 
(\q^1 \bar{\Pi} \bar{\g}^{I} \ep^2 + \q^2 \bar{\Pi} \bar{\g}^{I} \ep^1) 
\right ). \label{bterm}
\ee
This term does not in general vanish even 
given the relation (\ref{rel1}), along
with the conditions $\q^1 = \W \q^2$, $[\g^{r},\W] = 0$, 
$\del_{\s} x^r = 0$, and $\{ \g^{r'}, \W \} = 0$. 
 
This boundary term vanishes only when 
the fermionic variations in (\ref{z3}) 
are consistent with the fermion boundary conditions. 
So one should substitute $\ep^1 = \W \ep^2$
into the fermionic variations and demand
that they preserve $ \q^1 = \W \q^2$ on the boundary. Now
\bea
p^+ (\d \q^1 - \W \d \q^2){|} &=& 
- \left ( \W \sum_{r=1}^{p-1} \del_{\s} x^{r} \g^{+r} +  \W 
\sum_{r'=p}^{8} \del_{\t} x^{r'} \g^{+r'} \right ) \ep^2 {\Big |}
\label{gen} \\
&& + \half m \left ( \sum_{r=1}^{p-1} (\Pi + \W \Pi \W) \g^{+r} x^r 
+ \sum_{r'=p}^{8} (\Pi - \W \Pi \W) \g^{+r'} x^{r'} \right ) \ep^2 {\Big |}. 
\nn
\eea
This must vanish for any of the $Q^-$ supersymmetry to be preserved.
To evaluate this expression we have used only $[ \W, \g^{r} ] = \{
\W, \g^{r'} \} = 0$; the first two terms manifestly drop out
when $x^{r}$ is pure Neumann and $x^{r'}$ is pure Dirichlet, but
we will use the more general expression later for mixed 
boundary conditions.

For our case (i.e. $\Pi \W \Pi \W = -1$), (\ref{gen}) reduces to  
\be
\sum_{r' = p}^{8} \left( - \del_{\t} x_0^{r'}(\s,\t) \W 
+ m x_0^{r'}(\s,\t) \Pi  \right)  \g^{+r'} \ep^2 {|}. \label{c2}
\ee
Note that the Neumann directions
drop out of the condition. (\ref{c2}) can only be satisfied if
the transverse positions are all zero. Thus
we see that when the $D_-$ brane is located at the origin,
(\ref{c2}) vanishes and  
half of the dynamical supersymmetries are preserved, 
namely the combinations 
\be
q^{-} = \half (Q^{-1} + \bar{\W} Q^{-2}),
\ee
but when the brane is located at some other position 
all $Q^-$ supersymmetries are broken. This is exactly 
what we found in the probe analysis \cite{ST}.
We shall now show, however,
that there are eight fermionic symmetries, which are 
combinations of worldsheet and $Q^-$ supersymmetries,
that are preserved by the brane even when it is located 
away from the origin. So in effect the brane away from 
the origin is as supersymmetric as the brane 
at the origin. 

Recall that under the fermionic transformation,
\bea \label{qk}
p^+ \d \q^{1} & = & i d_k \Pi \ep^2_k  \f_k  + \ep^1_k \tilde{\f}_k; \\
p^+ \d \q^{2} & = & - i \Pi d_{k} \ep^1_{k} \tilde{\f}_{k}
+ \ep^2_k  \f_k, \nn \\
\d x^- &=& i (\q^1 \bar{\gamma}^- \d \q^1 
+ \q^2 \bar{\gamma}^- \d \q^2), \nn
\eea
where $k$ is a complex parameter, the worldsheet Lagrangian
transforms as a total $\s$-derivative (since fermionic
variations satisfy the fermionic field equations). 
When $k$ is an integer, these are just the closed string 
worldsheet symmetries discussed in (\ref{Q+n}) and 
when the relation (\ref{opQ+}) is imposed these
become the stringy extensions of $Q^+$ transformations 
discussed above. For general $k$, however, the boundary 
terms are non-vanishing, and (\ref{qk}) does not lead to a 
conserved charge.
 
Now, since $x_0^{r'}(\s,\t)$ satisfies the bosonic field 
equation, it can always be expanded as
\be
x_0^{r'}(\s,\t) = \sum_k (c_k^{r'} \f_k + \tilde{c}_k^{r'} \tilde{\f}_k),
\ee
where the coefficients $c_k^{r'}$ and $\tilde{c}_k^{r'}$ 
are constants. For example, for a brane located at 
arbitrary transverse position, $x_0^{r'}$, 
$x_0^{r'}(\s,\t)$ is given in (\ref{dir2}) and can be rewritten
as
\be \label{x0c}
x_0^{r'}(\s,\t)=
{x_{0}^{r'}  \over e^{m \pi} + 1}
(\f_{im} + e^{m \pi} \f_{-im}).
\ee
Let us now try to find a (\ref{qk}) transformation such that 
combined with the $Q^-$ transformation it yields
a transformation which leaves the action invariant, including
boundary terms, and so leads to a conserved charge.
In other words, we want to choose $\ep_k^1$ and 
$\ep_k^2$ such that
\bea
0 = p^+ (\d\q^{1} - \W d \q^2){|} 
&=& \sum_{k} \left( \f_{k} \left (  c_{k}^{r'} (i \w_{k} \W + m \Pi) 
\g^{+r'} \ep^2 + (i d_k \Pi - \W) \ep^2_{k} \right ) {\Big |}\right.
\nn \\
&& \left. + \tilde{\f}_k \left( \tilde{c}_{k}^{r'} (i \w_{k} \W + m \Pi) 
\g^{+r'} \ep^2
+ (1 - i \W \Pi d_{k}) \ep_{k}^1 \right ) {\Big |} \right).
\eea
This equation has a solution provided that $d_k^2 \neq 1$ 
which implies $k \neq 0$:
\bea \label{qkpar}
\ep^1_k &=& - {1 \over d_k^2-1} \tilde{c}_{k}^{r'}
\left(i (2 \w_k -k) \W + (m+d_k \w_k) \Pi \right) \g^{+r'} \ep^2 \nn \\
\ep^2_k  &=& {1 \over d_k^2-1}  c_{k}^{r'}
\left(- k i + (m - d_k \w_k) \Pi \W  \right) \g^{+r'} \ep^2
\eea
It follows that in all cases in which $x_0^{r'}(\s, \t)$ has an 
expansion that does not involve the $k=0$ modes, the
D-brane preserves  eight fermionic symmetries, 
namely the combination  
of $Q^-$ and the special worldsheet transformations 
(\ref{qk}) with the parameter given by (\ref{qkpar}).
$k=0$ is a special case in that the
worldsheet symmetry $Q_{0}$ is precisely the spacetime kinematical
supersymmetry and the terms in $x_{0}^{r'}$ involve
$\sin(m\t)$ and $\cos(m\t)$.  This case is thus relevant 
for D-branes translated away from the origin
using the broken translational or rotational symmetries.
It is convenient to
discuss these in detail separately in the next section.

In the specific case of an $D_-$ brane located at constant position 
$x_0^{r'}$,  $x_0^{r'}(\s)$ is given in (\ref{x0c}) 
and (\ref{qkpar}) becomes,
\be
\e_{im }^2 = - \half {m \over e^{m \pi +1}} (1+\Pi \W) \g^{+r'} \ep^2, \qquad
\e_{-im }^2 = \half {m e ^{m \pi} \over e^{m \pi +1}} 
(1-\Pi \W) \g^{+r'} \ep^2,
\ee
The Noether charge associated with this symmetry is
\bea
\hat{Q}^{-} &=&  \frac{1}{\pi} \int^{\pi}_{0} d \s \left (
\sum_{r=1}^{p-1} (\del_{\t} x^{r} \bar{\g}^{r} \q_{N}
- \del_{\s} x^r \bar{\g}^{r} \q_{D} + m x^{r} \bar{\g}^{r} \W \Pi 
\q_{N}) \right . \\
&& \left . \sum_{r'=p}^{8} (\del_{\t} (x^{r'} - x_0^{r'}) 
\bar{\g}^{r'} \q_{D}
- \del_{\s} (x^{r'} - x_0^{r'}) \bar{\g}^{r'} \q_{N} 
- m (x^{r'} - x_0^{r'}) \bar{\g}^{r'} \W \Pi \q_{D}) \right ). \nn 
\eea
where $\q_N$ and $\q_D$ are defined in appendix D.
This charge when realised in terms of modes reproduces precisely $q^-$.

The result however holds much more generally: branes
with arbitrary boundary conditions $x^{r'}(x^+)$ preserve
one half of the closed string kinematical supersymmetries
as well as eight additional dynamical supersymmetries
which descend from a combination of closed string and
worldsheet symmetries. 

\subsubsection{$D_+$ branes}

We have seen in the previous section that the 
conditions for having unbroken kinematical supersymmetry
$Q^+$ are (\ref{c1}). 
For the $D_+$ branes the second condition is not satisfied
and all of the $Q^{+}$ supersymmetries are necessarily broken. 
However, this
symmetry is replaced by another fermionic symmetry. 
As we have discussed, the $Q^+$ symmetry is a particular
case of a worldsheet symmetry and the latter originate 
from the fact that the worldsheet action is quadratic in the fields.
This is true for the $D_+$ branes as well and it turns out that
these branes possess kinematical supersymmetries too.
 
The new kinematical supersymmetry rules for $n \neq 0$ are still 
given by (\ref{Q+n}) but the parameters now satisfy
\be
\ep_n^1 = c_{n}^2 (\W (1 - d_{n}^2) - 2 i d_{n} \Pi) \ep_n^2
\ee
where $m d_{n} = (\w_{n} - n)$, $c_n=(1+d_n^2)^{-1/2}$, 
$\w_{n} = {\rm{sgn}}(n) \sqrt{n^2 + m^2}$ and $n$ is an 
integer. The corresponding Noether charges can be obtained from (\ref{Qn1})
and (\ref{Qn2}) by combining
$(Q_{n}^1 + c_{n}^2 (\bar{\W} (1 - d_{n}^2) - 2 i d_{n} \bar{\Pi}) 
Q_{n}^2)$. 

There are some special cases as usual. Firstly, we cannot find
a solution for $n=0$, which reflects the total breaking of
the closed string symmetry $Q^+$. However, there is a solution for
$k = \pm i m$:
\bea \label{hatQ}
\d \q^1 &=& \caP_{+} \ep^{+} e^{m \s} + \caP_{-} \ep^{-} e^{-m \s}; \\
\d \q^2 &=& \Pi \caP_{+} \ep^{+} e^{m \s} - \Pi \caP_{-} \ep^{-} 
e^{-m \s}; \nn \\
p^{+} \d x^{-} &=& i \left ( \q^1 \bar{\g}^{-} \d \q^1 
+ \q^2 \bar{\g}^{-} \d \q^2 \right ), \nn
\eea
where
\be
\caP_{\pm} = \half ( 1 \pm \W \Pi).
\ee
It is these transformations that effectively 
replace the closed string kinematical supersymmetries 
$Q^+$. Notice that the transformations are similar to (\ref{z2}) 
in that they only 
depend on one worldsheet coordinate but in this case it is
the $\s$ coordinate rather the $\t$ coordinate that features in the formulas.
Thus, unlike (\ref{z2}), the transformation 
rules (\ref{hatQ}) cannot be expressed 
entirely in terms of target space coordinates.
The corresponding Noether current is 
\bea
\hat{Q}^{+ \t} &=& {2  p^+ \over \pi} \bar{\g}^{-} 
\left ( e^{m\s} ( \caP_{+} \q^1 + \caP_{+} \Pi \q^2)
+  e^{-m \s} (\caP_{-} \q^1 - \caP_{-} \Pi \q^2) \right ); \label{qhat} \\
\hat{Q}^{+ \s} &=& {2 p^+ \over \pi} \bar{\g}^{-}
\left ( e^{m\s} ( \caP_{+} \q^1 - \caP_{+} \Pi \q^2)
+  e^{-m \s} (\caP_{-} \q^1 + \caP_{-} \Pi \q^2) \right ). \nn
\eea
As we shall show in \cite{ST3}, the supercharge $\hat{Q}^{+}$ is proportional 
to the $D_+$ brane fermionic zero modes exactly as the supercharge $Q^+$
is for the $D_-$ branes. 

We should comment here that although $\hat{Q}^+$ is just one
of the infinite set of symmetries $Q_n$ it does play a distinguished
role. This will become apparent when we discuss the superalgebra
and spectrum \cite{ST3}. The superalgebra of the preserved 
closed string supercharges plus $\hat{Q}^+$ closes
(the introduction of $Q_n$ leads to an infinite extension).
$\hat{Q}^+$ commutes with the Hamiltonian
so physical states form corresponding supermultiplets.
In the $\mu \rightarrow 0$ limit, it becomes precisely
a target space supercharge. Although both $\hat{Q}^+$ and $Q_n$ act on the 
spectrum, the former being realised in terms of zero modes
is exceptional.

\bigskip

Let us now consider the dynamical supersymmetries.
For $D_+$ branes (\ref{gen}) reduces to 
\be
\half m \sum_{r=1}^{p-1} ( \Pi + \W \Pi \W) 
\g^{+r} x^{r} \ep^2 {|}= m \sum_{r=1}^{p-1} \Pi \g^{+r} x^r 
\ep^2 {|}. \label{c3}
\ee
where we consider standard Dirichlet 
boundary conditions, i.e. $\pa_\tau x_0^{r'}(\s, \t)=0$.
If there are no Neumann directions amongst the $x^{I}$, 
i.e. for a D1-brane, (\ref{c3}) vanishes and  half
of the $Q^-$ are preserved, the combinations $(Q^{-1} + \bar{\W} 
Q^{-2})$, regardless of transverse location. 
For all other possibilities the $Q^-$ supersymmetry is broken and the
boundary term in (\ref{bterm}) does not vanish. 

One may hope that similar considerations as in the 
previous cases would allow us to find eight additional fermionic 
symmetries. In this case, however, the violation
of the symmetry in (\ref{c3}) involves the restriction of the Neumann
coordinates to the boundary which is unknown. 
In contrast, in all previous cases the violating terms
were involving terms fully determined by the boundary conditions.
We conclude that all $D_+$ branes but the $D1$ brane do not preserve 
any dynamical supersymmetry.

\subsubsection{Symmetry-related D-branes}

As discussed in section \ref{eqbr1}, one can act by 
broken spacetime generators on a static brane 
to obtain a new brane. 
For every symmetry of the static brane there is necessarily 
a corresponding symmetry of the 
translated brane and the algebra of the corresponding
charges is the same in both cases. 

Recall that for a D-brane translated using the background
Killing symmetries $P^{r'}$ and $J^{+r'}$ 
the corresponding Dirichlet boundary
condition is (\ref{rbc}), which couples the $x^+$ and $x^{r'}$
variations on the boundary as
\be \la{mbcz}
\d x^{r'}{|} = 
(- a^{r'} \mu \sin (\mu x^+) + b^{r'} \mu \cos (\mu x^+)) \d x^+{|}. 
\ee
In this section we will use $t'$ to denote an
ordinary Dirichlet direction and reserve $r'$ and $s'$ to 
denote directions in which there are 
modified Dirichlet boundary conditions (\ref{rbc}). 
Throughout the section when discussing conserved charges 
we assume that all constant Dirichlet positions are zero; if
any are constant but 
non-zero we simply replace the conserved symmetry we identify 
by the same symmetry combined with the appropriate
worldsheet symmetries as discussed in the previous section.

Given such boundary conditions, the conserved bosonic charges of
the D-brane certainly include $(P^+,P^{r},J^{+r},J^{rs},J^{t'u'})$,
following the same analysis as before. 
The other bosonic charges are more subtle, however. 
The Hamiltonian is not conserved;
to preserve the boundary condition, we must consider
a combination of $P^{-}$, $J^{+r'}$ and $P^{r'}$ transformations. That
the Hamiltonian by itself is not a good conserved charge follows
from the time dependence of the boundary conditions. The action
of the combined transformation
\be
\hat{P}^{-} = P^{-} + \sum_{r'} (\mu b^{r'} P^{r'} - \mu^2 a^{r'} J^{+r'}),
\ee
is   
\bea
\d x^{+} &=& \ep^{+}; \hsp
\d x^{r'} = - a^{r'} \mu \sin (\mu x^+) \ep^+ + 
\mu b^{r'} \cos (\mu x^+) \ep^+; \nn \\
\d x^{-} &=& \mu^2 \sum_{r'} (a^{r'} 
\cos (\mu x^+) x^{r'} + b^{r'} \sin (\mu x^+) x^{r'})
\ep^+,
\eea
which manifestly preserves (\ref{mbcz}). The other modified charges
are
\bea
\hat{J}^{r's'} &=& J^{r's'} - a^{r'} P^{s'} - \mu b^{r'} J^{+s'} 
+ a^{s'} P^{r'} + \mu b^{s'} J^{+r'}; \\
\hat{J}^{r't'} &=& J^{r't'} - a^{r'} P^{t'} - \mu b^{r'} J^{+t'}. \nn 
\eea

\bigskip

Now let us consider the supercharges. Consider
first the $D_-$ branes. The kinematical supercharges $q^+$
are unaffected by the time dependent 
boundary conditions; the same is true for the fermionic
charges $Q_n$. The combination of dynamical supersymmetry
transformations with $\ep^1 = \W \ep^2$ preserves the time
dependent $x^{r'}$ boundary condition but following (\ref{gen})
\be
p^+ (\d \q^1 - \W \d \q^2){|}  
= m \g^{+} \sum_{r'} ( \sin(m \t) (b^{r'} \bar{\Pi} \bar{\g}^{r'} - 
a^{r'} \bar{\g}^{r'} \W) \ep^2  
+ \cos(m \t) (b^{r'} \bar{\g}^{r'} \W + a^{r'} 
\Pi \bar{\g}^{r'}) ) \ep^2.
\ee
On its own this does not preserve the fermion boundary condition
but it is a good symmetry when 
combined with $Q^{+}$ transformations with parameters
$\hat{\ep}^1$ and $\hat{\ep^2}$ such that 
\be
(\d \q^1 - \W \d \q^2){|} = (\cos(m \t)
+ \sin(m \t) \W \Pi) (\hat{\ep^2} + \W \hat{\ep}^1)
\ee
provided that 
\be
(\hat{\ep}^2 + \W \hat \ep^1)  = \mu \g^{+} \sum_{r'} 
(b^{r'} \bar{\W} \bar{\g}^{r'} - a^{r'}
\bar{\Pi} \bar{\g}^{r'}) \ep^2.
\ee
The corresponding Noether charge is then
\be \la{rcha}
\hat{q}^{-} = 
\half \left ( Q^{-1} + \bar{\W} Q^{-2} + \half \mu \sum_{r'} (b^{r'} \g^{r'+} 
- 
a^{r'} \bar{\g}^{r'} \W \Pi \g^{+}) (Q^{+2} + \bar{\W} Q^{+1}) \right ).
\ee
That the symmetry 
involves only this specific combination of $Q^+$ transformations
follows from the fact that kinematical transformations with 
$(\hat{\ep}^2 + \W \hat{\ep}^1) = 0$ automatically preserve
the fermion boundary conditions by themselves.
The charge (\ref{rcha})  has eight independent components in total. It
is interesting to prove this same result from a rather different
perspective using the kappa symmetry projections; we include
this in appendix E. 

Thus the brane preserves in total one half of the closed string
supersymmetries, as indeed it must, being related by symmetry to
the half supersymmetric branes in the previous subsection. 
Under the symmetry transformation, the preserved dynamical charge is shifted
by a term proportional to the kinematical charge. This is to be expected
since from the spacetime superalgebra 
the commutator of the translation and $Q^-$ is proportional to $Q^+$. 
We will make this statement more precise in \cite{ST3} when
we discuss the embedding of the symmetry-related brane algebra into
the closed string algebra.

There is one special case where these preserved dynamical charges
are actually the same ones as for the static brane: this requires
\be
\sum_{r'} (b^{r'} \g^{r'} - a^{r'} \g^{r'} \W \Pi) = 0.
\ee
If one considers a $(+,-,2,0)$ D3-brane,
and takes $a^{3} = r$, $b^{4}= r$, 
so the brane traces out a circle in the $(34)$ plane,
this condition is satisfied.
Such a brane arises from taking a specific Penrose limit of a 
giant graviton in $AdS_5 \times S^5$ \cite{Len} which are indeed 
known to be half supersymmetric rotating branes \cite{Itz, My1}. 
Penrose limits of giant gravitons were also discussed in 
\cite{ST,BHLN,TT}.

What is particularly interesting, however, is that we now see
that all $D_-$ branes still preserve one half of the closed string 
supersymmetries when they rotate in any transverse plane. 
This suggests that there are many more half supersymmetric rotating branes
in AdS than were previously known. Furthermore, translated
or boosted branes (by which we mean 
those with $b^{r'}=0$ and $a^{r'} = 0$, respectively) may also originate from 
as yet unknown time dependent embeddings in AdS which are half 
supersymmetric. Following the same approach as here, 
such branes could
be obtained from the half supersymmetric 
static embeddings given in \cite{ST} by the
action of the symmetries of the AdS background which are
broken by the static brane.  

\bigskip

Now consider the $D_+$ branes. As for the static branes
closed string kinematical
supersymmetries are completely broken by the fermion 
boundary conditions,
but the branes possess different kinematical supersymmetries instead. 
The preserved worldsheet fermionic symmetries $Q_n$ are also
the same as for the static branes.  
Moving now to the dynamical supersymmetry, we observe that 
for the symmetry-related D1-branes the preserved Noether charges are
the combinations 
\be
\hat{q}^{-} = \half (Q^{-1} + \bar{\W} Q^{-2})
+ \qu \mu \sum_{r'} \left ( (b^{r'} \g^{r'+} - a^{r'} 
\bar{\g}^{r'} \W \Pi \g^{+}) Q^{+2}
+ (b^{r'} \g^{r'+} \bar{\W}  + a^{r'} 
\g^{r'+} \bar{\Pi}) Q^{+1}  \right ).
\ee
Thus a D1-brane carrying angular momentum in a transverse plane still
preserves one half of the dynamical supersymmetry.

This result fits in neatly with \cite{ST}, where it was
noted that $(+,1,0)$ D1-branes preserve one quarter of the closed
string supersymmetries. It
was then argued that these should arise from the Penrose limits of
rotating D-strings in $AdS_5 \times S^5$. A distinct Penrose
limit of these same rotating D-strings gives the 
$(+,-,0,0)$ branes carrying angular momentum in a single
transverse plane, which we have just shown also preserve one quarter
of the closed string supersymmetries (though let us emphasise that like
the static branes they 
do possess eight additional kinematical supersymmetries as well). 

\section{Conclusions}

We studied in this paper D-branes in the maximally supersymmetric 
plane wave background of IIB supergravity. We found both time-dependent 
and static branes. The branes fall into equivalence classes depending
on whether they are related by the action of target space
isometries. In particular, static branes localized at the origin
are in the same equivalence class as certain time-dependent branes.
The symmetries of all branes in the same equivalence
class are the same, but the relation of the conserved charges
with the conserved charges of the closed string is different
for different members of the equivalence class.  

Let us summarize the results for the supersymmetric static Dp-branes.
The D-branes fall into two categories: the $D_-$ branes
satisfying $\Pi \W \Pi \W = -1$, where $\W$ is related to the 
fermion boundary conditions (see (\ref{bc2})), and $D_+$ 
branes satisfying $\Pi \W \Pi \W = 1$. The $D_-$ branes
have worldvolume coordinates $(+,-,m,m \pm 2)$ and 
the $D_+$ are the remaining $(+,-,m,n)$ branes.

$D_-$ branes always preserve eight kinematical $Q^{+}$ supersymmetries
irrespective of location. They preserve eight dynamical 
supersymmetries $Q^{-}$ only if the brane is located
at the origin, and away from the origin $Q^{-}$ is broken
but we find instead eight new dynamical supercharges, $\hat{Q}^-$.
The latter are a linear combination of $Q^-$ with a certain 
fermionic worldsheet symmetry. $\hat{Q}^-$ anticommute to the Hamiltonian
much the same way as $Q^{-}$ do for the branes at the origin.
Actually, as we will show in \cite{ST3}, $\hat{Q}^-$ evaluated on-shell 
is equal to $Q^-$ on-shell (but the latter is relevant for
branes at the origin and the former for branes away from the origin).
Furthermore, we have seen that a far stronger result also holds: 
$D_-$ branes with arbitrary time dependent Dirichlet boundary conditions
of the form $x^{r'}(x^+)$
will also preserve not just half of the closed string kinematical
supersymmetries but in addition eight dynamical supersymmetries,
the combination of closed string with worldsheet symmetries. 

In $D_+$ branes all of the
$Q^+$ supersymmetries are broken, but a different set of 
8 kinematical supersymmetries $\hat{Q}^+$ is present and one half of the $Q^-$ 
are preserved if and only if there are no Neumann directions transverse
to the lightcone, i.e. it is a D1-brane\footnote{We do not consider 
branes with worldvolume flux in this paper. We note that the probe analysis in
\cite{ST} yielded $D_+$ $(+,-,0,4)$ and $(+,-,4,0)$ branes that preserve 
1/2 of dynamical supersymmetry.}. Thus, all $D_+$ branes but the 
D1-brane preserve 8 supercharges, and the latter preserves 16 supercharges.
The number of supersymmetries
that directly originate from the closed string supersymmetries
agrees with the probe analysis \cite{ST} in all cases.

We have seen that the light-cone action admits an infinite
number of worldsheet symmetries. These symmetries should 
also be present before gauge fixing the $\k$-symmetry.
It would be interesting to understand this in detail,
since this issue is closely related to the question
of under what conditions a given sigma model is related to an 
integrable theory. For a related discussion we refer to \cite{MM}.

A motivation for this work was to gain intuition about string theory
on curved backgrounds with RR fluxes. 
One of the main messages of this paper is that 
even the question of which are the possible supersymmetric states of 
string theory on such backgrounds (supersymmetric branes in our case)
is subtle. We have seen that tree-level open string theory yields
more supersymmetric branes than the probe analysis. The additional
supersymmetries were based on specific properties of the 
plane wave background. An important issue is whether the extra symmetries 
are respected by interactions. Provided that the answer is affirmative, 
consistency requires that the worldvolume
theory of D-branes in the plane wave background exhibits more supersymmetries
than the generic kappa-symmetry analysis yields
(but if string interactions invalidate the extra symmmetries
only the quadratic part of the worldvolume theory that captures
the spectrum of low energy excitations should exhibit more
supersymmetries).
In other words, our results indicate that the generic analysis of
the supersymmetry of embeddings only yields a lower bound on 
the number of supersymmetries rather than the exact number.

Given that the worldsheet theory is solvable only for very specific
backgrounds but the probe analysis is possible more generally, 
it is important to understand under which conditions one needs to 
amend the standard kappa-symmetry analysis. More generally 
it would be interesting to understand which backgrounds exhibit  
similar symmetry restoration mechanisms.  
 
In this paper we have studied open strings in the light-cone gauge.
It would be interesting to extend the analysis to more covariant
gauges which would allow, amongst other things, the study  
of branes not accessible in light-cone gauge, such 
as the supersymmetric branes found in \cite{ST} where  
only one of the lightcone directions is a worldvolume coordinate.
A description of the closed string in the plane wave background 
where all target space symmetries are manifest 
was given in \cite{Berkovits:2002zv}
using the pure spinor formalism. Worldsheet superconformal
field theories associated with the plane wave (and other 
RR backgrounds) were presented in \cite{Berkovits:2002vn,Berkovits:2002rd}
using the $U(4)$ hybrid formalism. It would be interesting to 
analyze D-branes using these formalisms, particularly
to understand the relation between the
worldsheet symmetries we find here and worldsheet superconformal
symmetries. Furthermore, quantum properties of the D-branes 
may be more transparent in the $U(4)$ formalism. 
 
In \cite{ST3} we continue the analysis started here by computing the 
corresponding superalgebras and the spectra of branes in all 
cases and comparing with the spectrum of small fluctuations 
around the corresponding D-brane emdedding in the DBI. 
Finally let us comment
that finding the corresponding boundary states
and checking open/closed duality for the branes
found here which are not included in the analysis
of \cite{BGG} is an important issue which we will also address
elsewhere.

\section*{Acknowledgments} 

We thank Universit\`{a} di Roma ``Tor Vergata'' for warm
hospitality at the initial stages of this work.
KS would also like to thank the Isaac Newton Institute, the 
Amsterdam Summer Workshop, and the Aspen Center for Physics   
for hospitality during the completion of this work. 
This material is based upon work supported by the National Science
Foundation under Grant No. PHY-9802484.
Any opinions, findings, and conclusions or recommendations expressed in
this material are those of the authors and do not necessarily reflect
the views of the National Science Foundation.

\appendix

\section{Conventions}

We follow closely the conventions of \cite{Met}. The Dirac
matrices in ten dimensions $\G^{\mu}$ are decomposed in
terms of $16$-dimensional gamma matrices $\g^{\mu}$ such that
\be
\G^{\mu} = \pmatrix{ 0 & \g^{\mu} \cr
\bar{\g}^{\mu} & 0 }, 
\ee
where
\bea
\g^{\mu} \bar{\g}^{\nu} + \g^{\nu} \bar{\g}^{\mu} = 2 \eta^{\mu \nu},
\hsp \g^{\mu} = (\g^{\mu})^{\a\b}, \hsp \bar{\g}^{\mu} = \g^{\mu}_{\a\b}, \\
\g^{\mu} = (1, \g^{I}, \g^{9}), \hsp 
\bar{\g}^{\mu} = (-1,\g^{I},\g^9).
\eea
Here $(\a,\b)$ are $SO(9,1)$ spinor indices in chiral representation;
we use the Majorana representation for $\G$ such that $C = \G^0$, and
so all $\g^{\mu}$ are real and symmetric. We use the convention that
$\g^{\mu_1..\mu_k}$ are the antisymmetrised product of $k$ gamma matrices
with unit weight. 

We assume the normalisation $\g^{0} \bar{\g}^1... \g^{8} \bar{\g}^9 = 1$,
so that
\be
\G_{11} = \G^{0}... \G^9 = \pmatrix {1 & 0 \cr 0 & -1}.
\ee
We define
\be
\Pi^{\a}_{\sp \b} \equiv (\g^1 \bar{\g}^2 \g^3 \bar{\g}^4)^{\a}_{\sp \b}, \hsp
(\Pi')^{\a}_{\sp \b} \equiv (\g^5 \bar{\g}^6 \g^7 \bar{\g}^8)^{\a}_{\sp \b}, 
\hsp
\g^{0} \bar{\g}^{9} = \g^{+-}.
\ee
Other useful relations are
\bea
\g^{+-} \Pi \Pi' = 1, \hsp (\g^{+-})^2 = \Pi^2 = (\Pi')^2 = 1, \\
\g^{+-} \g^{\pm} = \pm \g^{\pm}, \hsp 
\bar{\g}^{\pm} \g^{+-} = \mp \bar{\g}^{\pm}, \hsp
\g^{+} \bar{\g}^{+} = \g^- \bar{\g}^{-} = 0.
\eea
Abbreviated notations are used, in which the spinor indices
are indicated by the positioning of matrices; frequently
used expressions in the text include
\bea
&& \W Q \equiv \bar{\W}^{\sp \b}_{\a} Q_{\b}; \hsp
Q \W \equiv Q_{\a} \W^{\a}_{\sp \b}; \hsp
\W^{t} Q \equiv (\bar{\W}^{t})_{\a}^{\sp \b} Q_{\b}; \hsp
Q \W^t \equiv Q_{\a} (\W^{t})^{\a}_{\sp \b}; \\
&& \W^{\a}_{\sp \b} = (\g^{i_1})^{\a\g}.... (\bar{\g}^{i_q})_{\d \b}; \hsp
(\W^{t})^{\a}_{\sp \b} = ({\g}^{i_q})^{\a \g} ... 
(\bar{\g}^{i_1})_{\d \b}; \\
&& \bar{\W}_{\a}^{\sp \b} = 
(\bar{\g}^{i_1})_{\a\g}.... ({\g}^{i_q})^{\d \b}; \hsp
(\bar{\W^{t}})_{\a}^{\sp \b} = 
(\bar{\g}^{i_q})_{\a\g}.... ({\g}^{i_1})^{\d \b}.
\eea

\section{Symmetry currents and closed string superalgebra} \la{calg}

Here we give the explicit forms of the bosonic symmetry
currents in lightcone gauge; the fermion currents
are given in the main text. The momenta are
\bea
{\cal P}^{+ \t} &=& p^+; \hsp P^{+\s} = 0; \\
{\cal P}^{I \t} &=& 
(\cos (m \t) \del_{\t} x^{I} + m \sin (m \t) x^{I}); \\
{\cal P}^{I \s} &=& - \cos (m \t) \del_{\s} x^{I}.
\eea
The rotation currents are 
\bea
{\cal J}^{+I \t} &=& ( \mu^{-1} \sin (m \t) 
\del_{\t} x^{I} - p^+ x^{I} \cos (m \t) ); \\ 
{\cal J}^{+I \s} &=&  - \mu^{-1} \sin (m \t) 
\del_{\s} x^{I}; \\
{\cal J}^{ij \t} &=& \left (x^{i} \del_{\t} x^{j}
 - x^{j} \del_{\t} x^{i} - \half i  p^+ (
{\q}^1 \g^{-ij} \q^1 +  \q^2 \g^{-ij} \q^2 ) \right ); \\
{\cal J}^{ij \s} &=& - \left (x^{i} \del_{\s} x^{j}
 - x^{j} \del_{\s} x^{i} + \half i p^+
({\q}^1 \g^{-ij} \q^1 - \q^2 \g^{-ij} \q^2 ) \right ),
\eea
with a corresponding expression for ${\cal J}^{i'j'}$. The
remaining charge is the lightcone Hamiltonian.
\bea
{\cal P}^{- \t} &=& (\del_{\t} x^{-} + i ({\q}^1 
\bar{\g}^{-} \del_{+} \q^1 + \q^2 \bar{\g}^{-} \del_{-} {\q}^2)
- \mu m (x^{I})^2  - 4 i m \q^1 \bar{\g}^{-} \Pi \q^2); \\
{\cal P}^{- \s} &=& - (\del_{\s} x^{-} + i ({\q}^1 
\bar{\g}^{-} \del_{+} \q^1 - \q^2 \bar{\g}^{-} \del_{-} {\q}^2)).
\eea
Using the Virasoro constraint, and the fermion field equations,
we find that the onshell Hamiltonian density is
\be
{\cal H} = - {\cal P}^{- \t} = \frac{1}{2p^+} \left ( (\del_{\t} x^I)^2 + 
(\del_{\s} x^{I})^2 + m^2 (x^I)^2 \right) 
+ i (\q^1 \bar{\g}^{-} \del_{\t} \q^1 
+ \q^2 \bar{\g}^{-} \del_{\t} \q^2) .
\ee
Finally, let us remark that onshell ${\cal P}^{- \s} = - \del_{\s} x^-$. 

The symmetry superalgebra of the pp-wave background is as follows.
The commutators of the bosonic generators are\footnote{Notice
our rotational generators differ from the ones in \cite{Met} 
by a factor of $i$.}
\bea 
[P^{-}, P^{I}] &=& i \mu^2 J^{+ I}, \hsp
[P^{I},J^{+J}] = i \d^{IJ} P^{+}, \hsp
[P^{-}, J^{+I}] = - i P^{I}, \\
\left [ P^{i}, J^{jk} \right ] &=& -i (\d^{ij} P^{k} - \d^{ik} P^{j}), \hsp
[ P^{i'}, J^{j'k'} ] = -i (\d^{i'j'} P^{k'} - \d^{i'k'} P^{j'}), 
\nn \\
\left [ J^{+i}, J^{jk} \right ] &=& -i (\d^{ij} J^{+ k} - \d^{ik} J^{+j}), \hsp
[ J^{+i'}, J^{j'k'} ] = -i (\d^{i'j'} J^{+ k'} - \d^{i'k'} J^{+j'}), \nn \\
\left [ J^{ij}, J^{kl} \right ] &=& -i (\d^{jk} J^{il} + ...), \hsp
[ J^{i'j'}, J^{k'l'} ] = -i (\d^{j'k'} J^{i'l'} + ...), \nn 
\eea
where the ellipses denote permutations,
whilst the commutation relations between the even and odd generators are
\bea
\left [J^{ij},Q^{\pm} \right ] &=& -\half i Q^{\pm} (\g^{ij}), \hsp
[J^{i'j'},Q^{\pm} ] = -\half i Q^{\pm} (\g^{i'j'}), \nn \\
\left [J^{+I},Q^{-} \right ] &=& - \half i Q^{+} (\g^{+I}),  \\
\left [P^{I},Q^{-} \right ] &=& \half \mu Q^{+} (\Pi  \g^{+I}), \hsp
[P^{-}, Q^{+}] = \mu Q^{+} \Pi , \nn 
\eea
and the anticommutation relations are
\bea
\{ Q^{+}, \bar{Q}^{+} \} &=& 2 P^{+} \bar{\g}^{-}, \nn \\
\{ Q^{+}, \bar{Q}^{-} \} &=& (\bar{\g}^{-}\g^{+}\bar{\g}^{I}) P^{I} 
- i \mu (\bar{\g}^{-}\g^{+}\bar{\g}^{I} \Pi) J^{+I}, 
\\
\{ Q^{-}, \bar{Q}^{+} \} &=& (\bar{\g}^{+}\g^{-}\bar{\g}^{I}) P^{I} 
- i \mu (\bar{\g}^{+}\g^{-}\bar{\g}^{I} \Pi) J^{+I}, 
\nn \\
\{ Q^{-} , \bar{Q}^{-} \} &=& 2 \bar{\g}^{+} P^{-}  + i \mu (\g^{+ij} \Pi)
J^{ij} + i \mu (\g^{+i'j'} \Pi') J^{i'j'}. \nn
\eea

\section{Mode expansions for closed strings} \label{mcl}

In our conventions the closed string mode expansions are given by
\bea
x^{I}(\s,\t) &=& \cos (m \t) x_{0}^I + m^{-1} 
\sin (m\t) p_{0}^{I} + i \sum_{n \neq 0} \w_{n}^{-1} (\a_n^{1 I} 
\td{\f}_n + \a_n^{2I} \f_n); \\
\q^1 (\s,\t) &=& 
\q^1_0 \cos (m\t) + \Pi {\q}^2_{0} \sin (m\t) 
+ \sum_{n \neq 0} c_n \left ( i d_{n}\Pi \q^2_n \phi_n
+ {\q}^1_{n} \td{\f}_{n} \right ) ; \label{fercl} \\
\q^2 (\s,\t) &=& {\q}^2_0 \cos (m\t) - \Pi {\q}^1_{0} \sin (m\t) 
+ \sum_{n \neq 0} c_n \left ( - i d_n \Pi {\q}^1_n \td{\f}_n
+ {\q}^2_{n} \f_{n} \right ).
\eea
After canonical quantization we get the following (anti)commutators
\bea 
[p_{0}^{I},x_{0}^{J}] = - i \d^{IJ}, \hsp
[\a_{m}^{\ca I},\a_{n}^{\caj J}] = \half \w_{m} \d_{n+m,0} 
\d^{\ca \caj} \d^{IJ}, \\
\{ \q_{0}^{\ca}, \q_{0}^{\caj} \} = \frac{1}{4 p^+} (\g^+) \d^{\ca \caj},
\hsp
\{ \q_{m}^{\ca}, \q_{n}^{\caj} \} = \frac{1}{4 p^+} (\g^+) 
\d^{\ca \caj} \d_{m+n,0}, \label{clcm}
\eea
where $\ca = 1,2$.
It is convenient to introduce creation and annihilation operators
\be 
a_{0}^{I} = \frac{1}{\sqrt{2m}} (p_0^{I} + i mx_{0}^{I}), \hsp
\bar{a}_{0}^{I} = \frac{1}{\sqrt{2m}} (p_0^{I} - i m x_{0}^I), \hsp
[\bar{a}_0^{I},a_{0}^J] = \d^{IJ}.
\ee
Expressed in terms of these modes the spacetime charges are
\bea
P^{+} = p^{+}, \hsp
P^{I} = p_{0}^{I}, \hsp
J^{+I} = - x_{0}^{I} p^+, \\
Q^{+1} = 2 p^{+} \bar{\g}^{-} \q_0^2, \hsp
Q^{+2} = - 2 p^{+} \bar{\g}^{-} \q_0^1.
\eea
Note that the complex $Q^{+}$ appearing in the closed string 
algebra is
\be
Q^+ = \frac{1}{\sqrt{2}} (iQ^{+1} - Q^{+2}) = 2 p^{+} \bar{\g}^{-} \q_0.
\ee 
The rotation charges are 
\bea
J^{IJ} &=& - i ( a_{0}^{I} \bar{a}_{0}^{J} - a_{0}^{J} \bar{a}_{0}^{I}
+ \half \sum_{\ca =1,2} \q_{0}^{\ca} \g^{- IJ} \q_{0}^{\ca})
\\
&&  - i \sum_{\ca = 1,2} \sum_{n > 0}
\left ( 2 \w_{n}^{-1} 
(\a_{-n}^{\ca I} {\a}_{n}^{\ca J} - \a_{-n}^{\ca J} \a^{\ca I}_{n}) 
+ p^{+} \q_{-n}^{\ca} \g^{-IJ} \q_{n}^{\ca} \right ). \nn 
\eea
The Hamiltonian is
\bea
H &=& \frac{1}{2p^+} (p_0^2 + m^2 x_{0}^2) + 2 i m 
(\q_0^1 \bar{\g}^{-} \Pi \q_{0}^2) \\
&& + \sum_{\ca = 1,2} \sum_{n \neq 0} (\frac{1}{p^+} 
\a_{-n}^{\ca I} \a_n^{\ca I} + \w_n \q_{-n}^{\ca} \bar{\g}^{-} 
\q^{\ca}_n). \nn
\eea
Finally the dynamical supercharges take the form
\bea
Q^{-1} &=& (2 p_{0}^{I} \bar{\g}^{I} \q_0^{1} - 2m x_{0}^{I} 
\bar{\g}^{I} \Pi \q^2_0) \\ 
&& + \sum_{n > 0} \left ( 4 c_n \a^{I1}_{-n} 
\bar{\g}^{I} \q_n^{1} + \frac{2im}{\w_n c_n} \a_{-n}^{2I} 
\bar{\g}^{I} \Pi \q^{2}_n + h.c. \right ) \nn \\
Q^{-2} &=& (2 p_{0}^{I} \bar{\g}^{I} \q_0^{2} + 2m x_{0}^{I} 
\bar{\g}^{I} \Pi \q^1_0) \nn \\
&& + \sum_{n > 0} \left ( 4 c_n \a^{I2}_{-n} 
\bar{\g}^{I} \q_n^{2} - \frac{2im}{\w_n c_n} \a_{-n}^{1I} 
\bar{\g}^{I} \Pi \q^{1}_n + h.c. \right ) \nn 
\eea
The complex $Q^-$ appearing in the algebra is
$(Q^{-1} + i Q^{-2})/\sqrt{2}$.

\section{Poisson-Dirac brackets for oscillators} \label{PD}

Each of the classical commutation relations is of the generic form
\be
[z^{m}(\s),z^{n}(\s)] = h^{mn} \d (\s,\s'),
\ee
where $z^{m}$ denotes both scalars and fermions. To evaluate
the Poisson-Dirac brackets for the oscillators, we should use 
the following procedure. 

Firstly recall that given an orthonormal complete set of 
functions, $u_n(\s)$, over some domain,
the delta function can be represented as,
\be
\d(\s,\s') = \sum u_n(\s) u_n(\s')
\ee
In the domain $[0,\pi]$ there are several distinct sets of complete 
orthonormal functions:
\bea
u_0&=&{1 \over \sqrt{\pi}}, \qquad u_n = \sqrt{{2 \over \pi}} \cos n \s, 
\ n=1,2,... \\
v_n &=& \sqrt{{2 \over \pi}} \sin n \s,  \ n=1,2,...
\eea
These give rise to the following representations of the 
delta function
\bea
\d(\s,\s') &=& \frac{1}{\pi} + \sum^{\infty}_{n=1} \frac{2}{\pi} 
\cos (n \s) \cos (n \s')=
\sum_{-\infty < n < \infty} \frac{1}{\pi} 
\cos (n \s) \cos (n \s'), \\
&=& \sum^{\infty}_{n=1} \frac{2}{\pi} 
\sin (n \s) \sin (n \s') = \sum_{-\infty < n < \infty} \frac{1}{\pi} 
\sin (n \s) \sin (n \s'). \nn 
\eea
Now to evaluate commutation relations for fields which
are already expanded in one of these bases is
straightforward. For example, scalars satisfying Neumann
or Dirichlet boundary conditions (\ref{neu}) and (\ref{dir})
are expanded in terms of cosines and sines respectively. Thus
we can substitute the delta function Fourier expansions 
directly into their commutation relations (\ref{com}) to determine
the commutation relations for the modes. This gives the
first terms in (\ref{mode}). 

\subsection{$D_-$ branes}

The fermion mode expansions (\ref{fer}) and (\ref{fer2}) do
not however involve mutually orthogonal functions. This
reflects the fact that $\q^1$ and $\q^2$ do not
satisfy pure Neumann or pure Dirichlet boundary conditions
at the string endpoints. However, certain combinations
of $\q^1$ and $\q^2$ will satisfy such conditions. 

Let us consider first the case $ \W \Pi \W \Pi = -1$; then
by construction $(\q^1 - \W \q^2)$ satisfies a Dirichlet
condition at the endpoints whilst $(\q^1 + \W \q^2)$ satisfies
a Neumann condition at the endpoints. In fact using (\ref{fer}) we
find that
\bea
\q_{D} (\s, \t) &=& (\q^1 - \W \q^2)(\s, \t)
= 2 \sum_{n} c_{n} e^{-i \w_{n} \t} (d_{n} \Pi + i \W ) \sin(n \s) 
\q_n; \label{qdir} \\ 
\q_{N} (\s,\t) &=& (\q^1 + \W \q^2) (\s,\t) \\
&=& 2 (\q_{0} \cos (m \t) - \W \Pi \q_{0} \sin(m \t)) +
2 \sum_{n \neq 0} c_{n} e^{-i \w_{n} \t} (i d _{n} \Pi + \W ) \cos(n \s) 
\q_n, \nn
\eea
which are manifestly both expansions in the orthogonal functions. 
Using the commutation relations (\ref{com}) we find that
\be
\{ \q_{D} (\s), \q_{D}(\s') \}_{D.B.} = \frac{i}{2p^+} 
\pi \g^+ \d(\s,\s'); \hsp
\{ \q_{N} (\s), \q_{N}(\s') \}_{D.B.} = \frac{i}{2 p^+} 
\pi \g^+ \d(\s,\s'), \label{qcom}
\ee
and substituting in appropriate mode expansions for the Dirac
delta function we are able to fix all fermionic oscillator 
anticommutators. This gives
\be
\{ \q^{\a}_{n},\q^{\b}_m \}_{D.B.} 
= \frac{1}{8 p^+} (\g^{+})^{\a\b} \d_{n+m,0}, \label{nzm}
\ee
which is in fact the same expression as in flat space. 

\subsection{$D_+$ branes}

Now let us consider the case $\W \Pi \W \Pi = 1$ which is
rather more subtle. By construction $(\q^1 - \W \q^2)$ again 
satisfies a pure Dirichlet boundary condition. Using the mode
expansions (\ref{fer2}) we find the explicit form to be
\be \la{bdir}
\q_{D} (\s, \t) = (\q^1 - \W \q^2)(\s, \t)
= 2 \sum_{n} c_{n} e^{-i \w_{n} \t} (d_{n} \Pi + i \W ) \sin(n \s) 
\q_n,
\ee
which is in fact the same as in (\ref{qdir}); it also satisfies
the commutation relation given in (\ref{qcom}). Thus for the non-zero
modes, the Poisson-Dirac brackets are the same as in the $\Pi \W \Pi \W
= -1$ case, namely (\ref{nzm}). 
To extract the brackets for the zero modes
is rather harder, because they drop out of both this Dirichlet
combination and of the natural Neumann combination which is 
\be
\q_{N} (\s,\t) = (\q^1 + \W \q^2)(\s, \t) 
- m \int^{\s} d \s' (\Pi \q^2 + \W \Pi \q^1) (\s',\t).
\ee
This is Neumann because using the field equations,
and the boundary conditions, its sigma derivative vanishes on the
boundary. Explicitly, 
\be \la{bneu}
\del_{\s} \q_{N} (\s,\t) = 2 \sum_{n} c_n \w_{n} e^{-i \w_n \t}
(i d_n \Pi - \W) \sin (n \s) \q_n.
\ee
To derive the zero mode commutation relations
we should work directly with the fields $\q^1, \q^2$ which are
not expanded in terms of orthogonal functions. 

Explicitly evaluating the anticommutators using the mode expansions
(\ref{fer2}) we find
\bea
\{ \q^1(\s, \t), \q^1(\s',\t) \}_{D.B.} &=& \frac{i}{4 p^+} \pi (\g^+)
\d(\s,\s') 
\label{xcom}\\
&=& \{ \caP_{+} \q^{+}_{0}, \caP_{+} \q^{+}_{0} \} e^{m (\s + \s')} 
+ \{ \caP_{+} \q^{+}_{0}, \caP_{-} \q^{-}_{0} \} e^{m (\s - \s')} \nn \\
&& + \{ \caP_{-} \q^{-}_{0}, \caP_{+} \q^{+}_{0} \} e^{-m (\s - \s')} 
+ \{ \caP_{-} \q^{-}_{0}, \caP_{-} \q^{-}_{0} \} e^{-m (\s + \s')}
\nn \\
&& + {1 \over p^+} \left( \sum_{n > 0} \qu i \g^{+} \cos n ( \s - \s') 
- i \sum_{n > 0} d_{n}^2 c_{n}^4 \g^{+} \cos n (\s + \s') \right. \nn \\
&& \left. + \half i \sum_{n > 0} d_n c_{n}^4 (1 - d_n^2) \g^{+} 
\W \Pi \sin n (\s + \s'). \right) \nn
\eea
In this expression we have substituted the brackets for the non-zero
modes. We have also set the anticommutators of the zero modes with 
non-zero modes to zero. Had these anticommutators being non-zero 
the right-hand side of the above expression would be time-dependent.  
The other anticommutators are 
\bea
\{ \q^1(\s, \t), \Pi \q^2(\s',\t) \}_{D.B.} &=& 0 \label{xcom3} \\
&=&  \{ \caP_{+} \q^{+}_{0}, \caP_{+} \q^{+}_{0} \} e^{m (\s + \s')} 
- \{ \caP_{+} \q^{+}_{0}, \caP_{-} \q^{-}_{0} \} e^{m (\s - \s')} \nn \\
&& + \{ \caP_{-} \q^{-}_{0}, \caP_{+} \q^{+}_{0} \} e^{-m (\s - \s')} 
- \{ \caP_{-} \q^{-}_{0}, \caP_{-} \q^{-}_{0} \} e^{-m (\s + \s')} 
\nn           \\ 
&& + {1 \over p^+} 
\left( i \sum_{n > 0} c_{n}^4 (1 - d_n^2)^2 \W \Pi \g^{+} 
\cos n (\s + \s') \right. \nn \\
&& \left. + \half i \sum_{n > 0} d_n c_{n}^4 (1 - d_n^2) \g^{+} 
\sin n (\s + \s') \right), \nn
\eea
and 
\bea
\{ \Pi \q^2(\s, \t), \Pi \q^2(\s',\t) \}_{D.B.} &=& 
\frac{i}{4 p^+} \pi (\g^+) \d(\s,\s') \label{xcom2} \\
&=& \{ \caP_{+} \q^{+}_{0}, \caP_{+} \q^{+}_{0} \} e^{m (\s + \s')} 
- \{ \caP_{+} \q^{+}_{0}, \caP_{-} \q^{-}_{0} \} e^{m (\s - \s')} \nn \\
&& - \{ \caP_{-} \q^{-}_{0}, \caP_{+} \q^{+}_{0} \} e^{-m (\s - \s')} 
+ \{ \caP_{-} \q^{-}_{0}, \caP_{-} \q^{-}_{0} \} e^{-m (\s + \s')}
\nn \\
&& + {1 \over p^+} \left( \sum_{n > 0} \qu i \g^{+} \cos n ( \s - \s') 
- i \sum_{n > 0} d_{n}^2 c_{n}^4 \g^{+} \cos n (\s + \s') \right. \nn \\
&& \left. + \half i \sum_{n > 0} d_n c_{n}^4 (1 - d_n^2) \g^{+} 
\W \Pi \sin n (\s + \s') \right), \nn
\eea
where again we have used the relations for non-zero modes. 
It is also convenient to insert factors of $\Pi$ into the brackets
as we have done. Note that
(\ref{xcom}) and (\ref{xcom2}) can be simplified using the identity
\be
\qu i \pi \d(\s,\s') = \eig i + \qu i \sum_{n > 0} \cos (n (\s -
\s')).
\ee
Now the Poisson-Dirac brackets for the fermionic zero modes
must satisfy all three expressions. Comparing (\ref{xcom}) and
(\ref{xcom2}) we find that 
\be
\{ \caP_{+} \q^{+}_{0}, \caP_{-} \q^{-}_{0} \}_{D.B.}  = 0.
\ee
This means that (\ref{xcom}) can be simplified to give
\be
\eig = a \caP_{+} e^{m (\s + \s')} + b \caP_{-} e^{- m (\s +\s')} 
- \sum_{n > 0} {m \over 4 (m^2 + n^2)} (m \cos n (\s + \s') 
-n \W \Pi \sin n (\s + \s')).  \label{xre1} 
\ee
Similarly (\ref{xcom3}) yields,
\be
0 = a \caP_{+} e^{m (\s + \s')} - b \caP_{-} e^{- m (\s +\s')} 
+ \sum_{n > 0} {n \over 4 (m^2 + n^2)} (m \sin n (\s + \s')
+n \W \Pi \cos n (\s + \s') ).  \label{xre2} 
\ee
where we anticipate that
\bea
\{ \caP_{+} \q^{+}_{0}, \caP_{+} \q^{+}_{0} \}_{D.B.}  = 
\frac{i}{p^+} a \caP_{+} \g^{+}. \\
\{ \caP_{-} \q^{-}_{0}, \caP_{-} \q^{-}_{0} \}_{D.B.}  = \frac{i}{p^+} 
b \caP_{-} \g^{+}. \nn
\eea
and we used the relations
\bea
d_n^2 c_n^4 &=& \frac{m^2}{4 (n^2 + m^2)}; \\
d_n (1 - d_n^2) c_n^4 &=& \frac{mn}{2 (n^2 + m^2)}, \nn \\
c_n^4 (1-d_n^2)^2 &=& {n^2 \over m^2 + n^2}
\eea
Notice that (\ref{xre2}) is equal to the $\s$ derivative 
of (\ref{xre1}), so we only need to solve the latter.
Since $a$ and $b$ are each multiplied by orthogonal projection
operators, we can obtain two decoupled equations by 
multiplying (\ref{xre1}) by $\caP_{+}$ and $\caP_{-}$,
respectively. The result is 
\bea
\eig &=& a e^{m (\s + \s')} - \sum_{n > 0} \frac{m}{4 (n^2+m^2)} 
( m \cos n (\s + \s') - n \sin n (\s + \s') ), \label{aeq} \\
\eig &=& b e^{-m (\s + \s')} - \sum_{n > 0} \frac{m}{4 (n^2+m^2)} 
( m \cos n (\s + \s') + n \sin n (\s + \s') ), \nn
\eea
Notice that the equation for $b$ is identical to the equation
for $a$ but with $m \to -m$. Thus, it is sufficient to solve 
(\ref{aeq}).

To solve (\ref{aeq}) we note that the variable $\s+\s'$
takes values in $[0, 2 \pi]$. The Fourier transform
of the exponential is given by
\be
e^{m (\s + \s')} = {1 \over \pi} (e^{2 \pi m} -1)
\left({1 \over 2 m} + \sum_{n > 0} \frac{m}{n^2+m^2} 
( m \cos n (\s + \s') - n \sin n (\s + \s') ) \right)
\ee
Inserting in (\ref{aeq}) we determine $a$,
\bea \label{ab}
a &=& \frac{\pi m}{4 (e^{2\pi m}  - 1)}; \\
b &=& \frac{\pi m e^{2 \pi m}}{4 (e^{2 \pi m} - 1)}. \nn  
\eea
where we used $b=a(-m)$. Note that these do reduce to the usual flat space 
anticommutators in the flat space ($m \rightarrow 0$) limit. 

Notice that the $(+,-,0,4)$ and $(+,-,4,0)$ branes are special
in that $\W \Pi = 1$, which implies $\caP_{-} = 0$ and
we only have the zero modes $\q_{0}^{+}$. The derivation
we presented still holds, but we need to set $\caP_{-} = 0$
and $\caP_{+} = 1$ in all formulas. In particular,
\be
\{ \q_{0}^{+}, \q_{0}^{+} \}_{D.B.} = \frac{i}{p^+} a \g^{+}.
\ee
with the same $a$ as in (\ref{ab}). For the corresponding
anti-branes with $\W \Pi = -1$, it is $\q_{0}^{-}$ and
$b$ which are non-zero.

\section{Kappa symmetry projections for time dependent branes}

In this appendix we use the kappa symmetry projection to
show the number of (closed string) supersymmetries preserved 
by rotating and time-dependent D3-branes.
 
In the plane wave background the curved space gamma matrices,
denoted here by $\G_{m}^{(c)}$, are related to the 32 dimensional 
flat space matrices given in appendix A as
\be
\G_{+}^{(c)} =  - \mu \G_{+} + 
\half \mu \sum_{I=1}^{8} (x^{I})^2 \G_{-};
\hsp 
\G_{-}^{(c)} = - \mu^{-1} \G_{-}; \hsp
\G_{I}^{(c)} = \G_{I}.
\ee
The Killing spinors of the background may conveniently be written as
\bea \label{kilsp}
\ep &=& (1 + i P) (\Psi_{R} + i \Psi_{I}); \\
P &=& - \half \G_{-} (\sum_{i=1}^{4} x^{i} \G_{i} \Pi + 
\sum_{i'=5}^{8} x^{i'} \G_{i'} \Pi'); \nn \\
\Psi_{R} &=& (\lambda_+ + \cos(\mu x^+) \lambda_{-} 
+ \Pi \sin (\mu x^+) \eta_{-}); \nn \\
\Psi_{I} &=& (\eta_{+} + \cos(\mu x^+) \eta_{-} 
- \Pi \sin (\mu x^+) \l_{-}), \nn 
\eea
where the 32 dimensional constant negative chirality 
spinors $\eta_{\pm}$ and $\lambda_{\pm}$ 
satisfy $\G_{\pm} \eta_{\pm} = \G_{\pm} \lambda_{\pm} = 0$.

The kappa symmetry projection associated with any brane
embedding is defined by \cite{kappa} 
\be
d^{p+1} \xi \Gamma = - e^{-\Phi} {\cal{L}}_{DBI}^{-1} 
e^{\cal F} \wedge X |_{vol}, \label{kap1} 
\ee
where $X = \bigoplus_{n} \G_{(2n)} K^n I$,
$|_{vol}$ indicates that one should pick the terms proportional 
to the volume form, and the operations $I$ and $K$ act on spinors
 as $I \psi = -i \psi$
and $K \psi = \psi^{\ast}$. ${\cal{L}}_{DBI}^{-1}$  is the value of the
DBI Lagrangian evaluated on the background. Here we have used the notation 
\be
\G_{(n)} = \frac{1}{n!} d\xi^{i_{n}} \wedge ... \wedge d\xi^{i_1} 
\G_{i_1...i_n},
\ee
where $\G_{i_1...i_n}$ is the pullback for the target space
gamma matrices 
\be
\G_{i_1...i_n} =\del_{i_1} X^{m_1} ... \del_{i_n} X^{m_n} 
\G_{m_1...m_n}^{(c)}.     
\ee
For the case in hand, a 
D3-brane whose worldvolume directions are $(x^+,x^-,x^1,x^2)$
and whose transverse positions are $x^{r'}(x^+)$, the kappa
symmetry projection is $\G \ep =  \ep$ where
\be \label{qr}
\G = - i (Q + R); \hsp 
Q = \G_{+-12}; \hsp R = \sum_{r'} \mu^{-1} \del_{+} x^{r'} \G_{-12 r'}.
\ee
and $\e$ is the killing spinor in (\ref{kilsp}) evaluate on the 
worldvolume of the D3-brane.
From the real and imaginary parts of the projection we get
two independent equations which are
\bea \la{k1}
Q \Psi_{R} + R \l_{+} - Q P \eta_{+} &=& - \Psi_{I} - P \eta_{+}; \\
Q \Psi_{I} + R \eta_{+} + Q P \l_{+} &=& \Psi_{R} - P \eta_{+}, \nn
\eea
where we have used $R \l_{-} = R \y_{-} = P \l_{-} = P \y_{-} = 0$. 
Inserting $\Psi_{R}$ and $\Psi_{I}$ from (\ref{kilsp}) we get
\bea \la{kzz}
(Q \l_+ + \eta_+)+
(Q \l_{-} + \eta_{-}) \cos(\mu x^+) + 
\Pi \sin(\mu x^+) (Q \y_{-} - \l_{-}) +
(R + P) \l_{+} &=& Q P \y_{+}; \qquad \\
(Q \eta_+ - \l_+)+
(Q \eta_{-} - \l_{-}) \cos(\mu x^+) 
- \Pi \sin(\mu x^+) (Q \l_{-} + \y_{-})
+ (R + P) \eta_{+} &=& - Q P \l_{+} \nn
\eea
These equations should hold for any values of the worldvolume 
coordinates. This implies that the part that is independent of the 
worldvolume coordinates should vanish separately,
\be \la{kz}
Q \l_{+} = - \y_{+}.
\ee
This relates the real and imaginary parts of the
parts of the spinor satisfying $\G_{+} \ep  = 0$. 
The remaining equations can be satisfied by imposing
separately:
\be \la{k2}
Q \l_{-} = - \y_{-}; \hsp R Q \y_{+} = [Q, P] \y_{+}, \hsp \{Q, R\}=0.
\ee
The last equation holds automatically for the $R$ and $Q$ given in 
(\ref{qr}). The first one can always be satisfied
for non-zero $\l_{-}$ independently of $x^{r'}(x^+)$; 
thus any time dependent branes always preserve at
least one quarter of the closed string supersymmetries,
that is, half the $Q^+$. The second equation in (\ref{k2}) imposes
further constraints on $\eta_{+}$ and determines 
which closed string dynamical supercharges
are preserved. As noted in \cite{ST}, this equation is
satisfied automatically if the constant transverse positions
are at the origin whilst if any are not
we project out the $\eta_{+}$ and $\l_{+}$ and break
all the $Q^{-}$. 

The solution in (\ref{k2}) is not the most general solution
of (\ref{kzz}), however. $R$ and $P$ may depend on $x^+$ in 
such a way that these terms may be combined with the remaining
terms in (\ref{kzz}) rather than treated separately as is done in 
(\ref{k2}). This is exactly what happens for the
symmetry-related branes. Recall that for these branes 
the relevant boundary conditions are (\ref{rbc}).
For these boundary conditions, we still 
have eight independent solutions to (\ref{kzz}) satisfying 
\be \la{up}
Q \l_{-} = - \eta_{-}; \hsp \eta_{+} = 0.
\ee
These are the eight kinematical supercharges which are preserved
irrespective of the time dependent boundary conditions. 
Furthermore, with the boundary conditions (\ref{rbc})
there are eight additional 
solutions to (\ref{kzz}) (provided that the other transverse
scalars are zero) for which
\be \la{pres}
\eta_{-} + Q \l_- = \sum_{r'} (b^{r'}\G_{-r'} - a^{r'} 
\Pi \G_{-12r'}) \eta_{+}.
\ee
with (\ref{kz}) also satisfied. The total number of preserved
supersymmetries is thus sixteen as expected. Furthermore
the specific combination of supersymmetries preserved in (\ref{pres}) 
agrees with that found in the open string analysis.

\end{document}